\documentclass[prd, aps, twocolumn, showpacs, nofootinbib,superscriptaddress]{revtex4}

\usepackage{amsmath,amssymb,amsthm,wasysym}
\usepackage{graphicx}
\usepackage{psfrag}
\usepackage{epstopdf}
\usepackage{color}
\usepackage{mathrsfs}
\usepackage{comment}

\DeclareSymbolFontAlphabet{\mathrsfs}{rsfs}
\DeclareMathAlphabet{\mathcal}{OMS}{cmsy}{m}{n}
\newcommand{\scri}{\mathrsfs{I}}

\def\lm{{\ell m}}

\def\de{\partial}
\def\lm{{\ell m}}

\def\ii{{\rm i}}

\def\F{{\cal F}}

\def\S{{\cal S}}
\def\O{{\cal O}}

\def\R{{\cal R}}

\def\rcm{R_*}

\newcommand{\be}{\begin{equation}}
\newcommand{\ee}{\end{equation}}

\usepackage{float}
\definecolor{cyan}{rgb}{0,0.9,0.9}
\definecolor{orange}{rgb}{0.9,0.5,0}
\definecolor{magenta}{rgb}{1,0,1}
\definecolor{purple}{rgb}{0.8,0.4,0.8}

\begin{document}

\title{Binary black hole coalescence in the large-mass-ratio
  limit:\\ the hyperboloidal layer method and waveforms at null infinity}

\author{Sebastiano \surname{Bernuzzi}}

\affiliation{Theoretical Physics Institute, University of Jena,
  07743 Jena, Germany}

\author{Alessandro \surname{Nagar}}
\affiliation{Institut des Hautes Etudes Scientifiques, 91440
  Bures-sur-Yvette, France}

\author{An{\i}l \surname{Zengino$\mathrm{\breve{g}}$lu}} 
\affiliation{Theoretical Astrophysics, California Institute of
  Technology, Pasadena, California, USA} 

\begin{abstract}
  We compute and analyze the gravitational waveform emitted to future
  null infinity by a system of two black holes in the large mass ratio
  limit. We consider the transition from the quasi-adiabatic
  inspiral to plunge, merger, and ringdown. 
  The relative dynamics is driven by a leading order in the
  mass ratio, 5PN-resummed, effective-one-body (EOB), analytic 
  radiation reaction. 
  To compute the waveforms we solve the Regge-Wheeler-Zerilli 
  equations in the time-domain on
  a spacelike foliation which coincides with the standard
  Schwarzschild foliation in the region including the motion of the
  small black hole, and is globally hyperboloidal,
  allowing us to include future null infinity in the
  computational domain by compactification. This method is called
  the hyperboloidal layer method, and is discussed here for the first
  time in a study of the gravitational radiation emitted by black hole
  binaries.  
  We consider binaries characterized by five mass ratios,
  $\nu=10^{-2,-3,-4,-5,-6}$, that are primary targets of 
  space-based or third-generation gravitational wave detectors.
  We show significative phase differences between finite-radius
  and null-infinity waveforms. We test, in our context, the reliability of the
  extrapolation procedure routinely applied to numerical relativity waveforms.
  We present an updated calculation of the final and maximum gravitational recoil 
  imparted  to the merger remnant by the gravitational wave emission,
  $v^{\rm end}_{\rm kick}/(c\nu^2)= 0.04474 \pm 0.00007$ and $v^{\rm max}_{\rm kick}/(c\nu^2)=0.05248\pm 0.00008$. 
  As a self consistency test of the method, we show an excellent fractional
  agreement (even during the plunge) between the 5PN EOB-resummed mechanical angular 
  momentum loss and the gravitational wave angular momentum flux computed at null infinity. 
  New results concerning the radiation emitted from unstable circular orbits are also presented. 
  The high accuracy waveforms computed here 
  could be considered for the construction of template banks or 
  for calibrating analytic models such as the effective-one-body model. 

\end{abstract}


\maketitle

\section{Introduction }

Compact binaries with large mass ratios are primary targets 
for space-based detectors of gravitational waves (GWs), like 
the Laser Interferometer Space Antenna (LISA)~\cite{lisa, esa-lisa} 
(or the similar ESA-led mission), and for third-generation 
ground-based detectors, like the planned 
Einstein Telescope~\cite{Punturo:2010zz}.
For example, the quasi-adiabatic inspiral of extreme-mass-ratio (EMR) binaries, i.e.~of mass ratio 
$\nu\sim10^{-6}$, is interesting for LISA (see e.g.~\cite{Sathyaprakash:2009xs}), 
while the merger of intermediate-mass-ratio (IMR) binaries, $\nu\sim 10^{-2}-10^{-3}$, 
is in the band of sensitivity of the Einstein Telescope~\cite{Huerta:2010un}.   
The theoretical modelling of such sources is a difficult task 
since neither numerical relativity (NR) simulations 
(due to their computational cost~\cite{Lousto:2010ut,Sperhake:2011ik}),  
nor standard post-Newtonian (PN) techniques~\cite{Blanchet:2006zz} 
(due to the strong-field, high-velocity regime) can be applied. 

Black-hole perturbation theory is instead the natural tool to model 
large mass ratio binaries~\cite{Davis:1971gg,Davis:1972ud,Lousto:1996sx,
Martel:2001yf,Nagar:2006xv,Mino:2008at,Bernuzzi:2010ty,
Sundararajan:2010sr,Berti:2010ce,Mitsou:2010jv,Hadar:2011vj}. 
The relative dynamics of the binary is
described by the motion of a particle (representing the small 
black hole) in a fixed background, black-hole spacetime 
(representing the central, supermassive black hole).
The dynamics of the particle is driven away from geodesic motion 
by the action of radiation reaction through a long, quasi-adiabatic 
inspiral phase up to the nonadiabatic plunge into the black hole. 
For what concerns nonconservative (dissipative) effects only, 
they can be modeled either numerically, for example in the adiabatic approximation, 
(e.g. as in~\cite{Sundararajan:2007jg,Sundararajan:2008zm,Sundararajan:2010sr} and references therein) 
or analytically, using PN-resummed results (\`a la effective-one-body), going 
in fact beyond the adiabatic 
approximation~\cite{Nagar:2006xv,Damour:2007xr,Bernuzzi:2010ty,Bernuzzi:2010xj}.  
Gravitational self-force calculations~\cite{Barack:2009ey,Barack:2009ux,
Barack:2010tm,Blanchet:2010zd, Barack:2011ed,Poisson:2011nh} 
can provide corrections to the particle conservative and
nonconservative dynamics at next-to-leading/higher order in the mass 
(away from geodesic motion), although the field is not 
ready yet for waveform production.
Finally, a very promising (semi)-analytical approach to describe the 
binary dynamics and to produce waveform template banks 
(for {\it any} mass ratio, including EMR and IMR binaries) 
is the effective-one-body (EOB) model~\cite{Buonanno:1998gg,Buonanno:2000ef,
Damour:2001tu,Buonanno:2005xu,Damour:2008gu,Damour:2008qf,Damour:2009wj,
Barausse:2009xi,Nagar:2011fx,Barausse:2011ys}.
The EOB approach is intrinsically nonadiabatic and it is designed to take into
account both conservative and nonconservative back-reaction effects,
but requires the calibration of some flexibility parameters to account 
for (yet uncalculated) higher-order effects in the dynamics 
and waveforms~\cite{Damour:2009kr,Baiotti:2011am,Pan:2010hz,Pan:2011gk,Yunes:2009ef,
Yunes:2010zj,Bernuzzi:2010xj,Tiec:2011bk}. 

The most important output of these studies is the GW signal
which encodes the gauge-invariant information about the source as it 
should be seen by detectors. Gravitational waves are rigorously and 
unambiguously defined only at null infinity. Numerical computations, however,
are confined to finite grids. A theoretical problem is thus to model and 
to compute the waveforms at null infinity, as seen by a far-away idealized
observer.  

This problem is prominent especially in NR simulations.
When an asymptotically Cauchy foliation of the spacetime is employed,  
the waveforms are typically extracted on coordinate spheres at finite 
distances from the source. 
To compute waveforms at null infinity post-simulation techniques are applied. 
Extrapolation to infinite extraction radius~\cite{Boyle:2007ft,Boyle:2008ge,
Boyle:2009vi,Pollney:2009ut,Pollney:2009yz} 
proved to be sufficiently robust and accurate, 
though somehow delicate due to ambiguities 
introduced by the gauge dynamics and the choice of a fiducial background.
An unambiguous procedure based on the Cauchy-characteristic extraction
(CCE) method~\cite{Bishop:1996gt, Babiuc:2005pg, Babiuc:2008qy}
has recently been implemented~\cite{Reisswig:2009us,Reisswig:2009rx,Babiuc:2011qi} 
to extract waveforms from binary black hole mergers of comparable masses.
Although the set up of initial data for the characteristic evolution
is intricate~\cite{Bishop:2011iu}, the method successfully provides waveforms from binary black 
hole mergers at null infinity and permits to cross-check the 
standard extrapolation procedure.

An alternative approach that does not require post-processing is to 
employ spacelike surfaces that approach null infinity. 
Such surfaces are called hyperboloidal because their asymptotic 
behavior resembles that of standard hyperboloids in Minkowski spacetime~\cite{Friedrich83}.
Hyperboloidal foliations have already been considered in the early days of
numerical relativity and were expected to be suitable for studying gravitational 
radiation~\cite{Smarr:1977uf, Eardley:1978tr, Brill80, Gowdy81}. The hyperboloidal initial value problem for the Einstein equations has been analyzed
by Friedrich~\cite{Friedrich83,Friedrich86}. His conformally regular field
equations have been implemented numerically in certain test cases
(for reviews see~\cite{Frauendiener:2000mk,Husa:2002zc}).

More recently, alternative hyperboloidal formulations have been 
suggested~\cite{Zenginoglu:2008pw,Moncrief:2008ie,Bardeen:2011ip} that
do not exhibit explicit conformal regularity. 
The only successful numerical implementation of such a formalism is by 
Rinne in axisymmetry~\cite{Rinne:2009qx}.  It is an outstanding question 
whether this or a similar hyperboloidal approach will lead to generic numerical simulations of black hole spacetimes. 

While the numerical properties of the hyperboloidal method 
for Einstein equations is only poorly understood in the general case, 
the situation is much clearer in perturbation theory where the background is
given. There, the best numerical gauge is to fix the coordinate location of null
infinity (scri), as first discussed by Frauendiener in the context of conformally
regular field equations~\cite{Frauendiener98b}. 
Moncrief presented the first explicit construction of a hyperboloidal scri-fixing 
gauge for Minkowski spacetime~\cite{Moncrief00} (for numerical implementations 
see \cite{Fodor:2003yg, Fodor:2006ue,Bizon:2008zd}).
The application of the 
method in black hole spacetimes proved to be difficult~\cite{Gentle:2000aq,
Gowdy:2001ij, Schmidt02, Calabrese:2005rs, vanMeter:2006mv}, 
until the general construction of suitable hyperboloidal scri-fixing 
coordinates on asymptotically flat spacetimes has been 
presented~\cite{Zenginoglu:2007jw}. 
Since then, hyperboloidal scri-fixing coordinates have been employed in a 
rich variety of problems concerning black hole 
spacetimes~\cite{Zenginoglu:2008wc, Zenginoglu:2008uc, Zenginoglu:2009ey,
Zenginoglu:2009hd, Gonzalez:2009hn, Bizon:2010mp, Zenginoglu:2010zm,
Bernuzzi:2010xj, LoraClavijo:2010xc, Vega:2011wf, Racz:2011qu}.

In particular, hyperboloidal compactification has been applied  to solve 
in time-domain the homogeneous Regge-Wheeler-Zerilli (RWZ) 
equations~\cite{Regge:1957td,Zerilli:1970se,Sarbach:2001qq,
Martel:2005ir,Nagar:2005ea}  for metric perturbations of a Schwarzschild black
hole~\cite{Zenginoglu:2009ey}. This work showed the efficiency of hyperboloidal
compactification as applied to the RWZ equations and discovered that the
asymptotic formula relating the curvature perturbation $\psi_4$ to the
gravitational strain is invalid for the polynomially decaying solution even at large
distances used for standard waveform extraction, thereby emphasizing the 
importance of including null infinity in numerical studies of gravitational radiation.

The solution of the {\it inhomogeneous}  RWZ equations on
a hyperboloidal slicing of the Schwarzschild spacetime is discussed 
in this paper for the first time.
The presence of a compactly supported matter source, 
such as a point-particle~\cite{Davis:1972ud,Lousto:1996sx} 
or a test-fluid~\cite{Papadopoulos:2000gb,Nagar:2005cj,Nagar:2006eu,Nagar:2006xv},
implies modifications.
It may be desirable to use standard techniques in a compact domain including the 
central black hole and the matter dynamics. The hyperboloidal method shall then be restricted 
to the asymptotic domain only, so that standard coordinates for matter dynamics can be employed.
Such a restricted hyperboloidal compactification provides the idealized waveform at null infinity, 
avoids outer boundary conditions, and increases the efficiency of the numerical computation without 
changing the coordinate description of matter dynamics. 

A convenient technique to achieve this, called 
the {\it hyperboloidal layer method}, has been introduced in~\cite{Zenginoglu:2010cq}. 
A hyperboloidal layer is a compact radial shell in which the spacelike foliation approaches 
null infinity and the radial coordinate is compactifying. By properly attaching such a layer
to a standard  computational domain, one makes sure that outgoing waves are 
transported to null infinity and no outer boundary conditions are needed. 
An intuitive prescription for the construction of a suitable hyperboloidal layer, 
that we describe in Sec.~\ref{sec:layer}, is to require that the spherically outgoing null 
surfaces have the same representation in the layer coordinates 
as in the interior coordinates. Because the hyperboloidal layer is practically attached 
to an existing computational domain, only minimal modifications to current numerical
infrastructures are needed for its implementation.

In this paper we apply the hyperboloidal layer method to improve the 
quality of recently computed RWZ waveforms emitted by the coalescence of 
(circularized) black-hole binaries in the test-particle
limit~\cite{Bernuzzi:2010ty} (hereafter Paper~I)
(see also Refs.~\cite{Nagar:2006xv,Damour:2007xr,Bernuzzi:2010xj}).
The central new result of this paper is the computation of highly accurate
gravitational waveforms at future null infinity ($\scri^+$) with an efficient and robust method.
As in Paper I, the relative motion of the binary is driven by 
5PN-accurate, EOB-resummed~\cite{Damour:2008gu,Fujita:2010xj} 
analytical radiation reaction and we focus on the transition 
from quasi-adiabatic inspiral to plunge, merger, and ringdown.
To span the range between IMR and EMR, we consider five mass ratios, 
$\nu\equiv\mu/M=10^{-2,-3,-4,-5,-6}$, where $M$ is the mass of 
the central Schwarzschild black hole, and $\mu$ is the mass
of the small compact object approximated as a point particle.
We estimate the differences between waveforms extracted at $\scri^+$  
and waveforms extracted at finite radii, and we provide an updated 
estimate of the gravitational recoil previously computed 
from finite-radius waveforms in
Refs.~\cite{Bernuzzi:2010ty,Sundararajan:2010sr}.
The availability of $\scri^+$ waveforms also allows us to assess, in a well 
controllable setup, the accuracy of the extrapolation procedure that is
routinely applied to NR waveforms.

The new multipolar waveform extracted at $\scri^+$ 
presented here has already been used in Ref.~\cite{Bernuzzi:2010xj} 
(hereafter Paper~II) to obtain several results that are valuable 
for currently ongoing EOB/NR comparisons: 
(i) finite-distance effects are significant even at comparatively 
large extraction radii ($r\sim 1000M$);
(ii) the agreement between the EOB-resummed analytical multipolar 
waveform~\cite{Damour:2008gu,Fujita:2010xj} and the RWZ waveform 
improves when the latter is extracted at $\scri^+$; 
(iii) the tuning of next-to-quasi-circular corrections to the phase 
and amplitude of the EOB-resummed (multipolar) waveform
improves  its agreement with the RWZ waveform during the late-plunge 
and merger phase 
(See also Ref.~\cite{Pan:2011gk} for a similar tuning procedure 
applied to several black-hole binaries with comparable mass ratios.)

The paper is organized as follows. In Sec.~II we briefly recall the model for 
the relative dynamics of the binary. The construction of the hyperboloidal layer 
in Schwarzschild spacetime is carried out in Sec.~III. We discuss the RWZ
equations with and without the hyperboloidal layer in Sec.~IV.
Details of the numerical implementation are presented in Sec.~V.
Physical results are collected in Sec.~VI, which consists of the following parts. 
First, we assess the accuracy of our implementation in the case of stable 
circular orbits, and present new results for unstable circular orbits. 
We then focus on the gravitational waveforms emitted during the transition from 
the quasi-circular inspiral through plunge, merger, and ringdown, and we quantify 
the differences with finite-radius extraction. We discuss the performance 
of standard techniques to extrapolate the finite-radius waveform to infinite 
extraction radius. Concluding remarks are presented in Sec.~VI. 
In Appendix~\ref{app:conv} we present convergence tests of the code.
In Appendix~\ref{app:asymp} we summarize the relations between the RWZ
master functions and asymptotic observables.
We mainly use geometrized units with $G=c=1$.

\section{Relative dynamics} 
\label{sec:rd}

The relative dynamics of the binary is computed as in Paper I and II;
here we review a few elements that are relevant to our study.

The binary dynamics has a conservative part (Hamiltonian)
and a dissipative part (radiation-reaction force).
The conservative part  
is described by the $\nu\to 0$ limit of the EOB Hamiltonian 
(the Hamiltonian of a particle in Schwarzschild spacetime) 
with the following, dimensionless variables: the relative separation 
$r=R/M$, the orbital phase $\varphi$, the orbital 
angular momentum $p_\varphi = P_\varphi/(\mu M)$, and the 
orbital linear momentum $p_{r_*}=P_{r_*}/\mu$, canonically 
conjugate to the tortoise radial coordinate separation $r_*=r + 2 \ln(r/2-1)$.
The Schwarzschild metric in standard coordinates $(t,r)$ reads
\be
\label{eq:ss} g=-A\,dt^2 + A^{-1}\,dr^2+r^2\,d\sigma^2 \ ,  
\ee 
where $d\sigma^2$ is the standard metric on the unit sphere and $A\equiv1-2/r$. 
The Schwarzschild Hamiltonian per unit ($\mu$) mass is
\begin{equation}
\hat{H} = \sqrt{ A\left( 1+ \frac{p_{\varphi}^2}{r^2}
  \right)+p_{r_*}^2}\ .
\end{equation}
The expression for the analytically resummed mechanical angular momentum loss
(our radiation-reaction force), $\hat{\cal F}_\varphi$,
is accurate at first order in the mass ratio, $\O(\nu)$, and is computed from
the 5PN-accurate EOB-resummed waveform of 
Refs.~\cite{Damour:2007xr,Damour:2008gu,Bernuzzi:2010ty,Fujita:2010xj}.  
Following~\cite{Damour:2006tr,Nagar:2006xv,Damour:2007xr,Bernuzzi:2010ty}, 
we use 
\be
\label{eq:rr}
\hat{\F}_\varphi \equiv -\dfrac{32}{5}\nu \Omega^5 r^4 \hat{f}(v_\varphi),
\ee
where  $\Omega=d\varphi/dt$ is the orbital frequency, $v_\varphi = r\Omega$ 
is the azimuthal velocity, and $\hat{f}=F^{\ell_{\rm max}}/F^{\rm Newt}_{22}$ 
denotes the Newton-normalized ($\nu=0$) energy flux up to multipolar
order $\ell_{\rm max}$, analytically resummed according 
to Ref.~\cite{Damour:2007xr,Damour:2008gu}. The resummation procedure 
is based on a certain multiplicative decomposition of the circularized 
multipolar gravitational waveform. More precisely, for circular orbits, 
the energy flux is written as
\begin{align}
\label{eq:Flux}
F^{\ell_{\rm max}}&=\sum_{\ell =2}^{\ell_{\rm max}} \sum_{m=1}^{\ell}F_\lm \nonumber\\
&=\dfrac{1}{8\pi}\sum_{\ell =2}^{\ell_{\rm max}} \sum_{m=1}^{\ell} (m\Omega)^2|r h_{\ell m}|^2.
\end{align}
Above, $h_{\ell m}$ is the factorized waveform of~\cite{Damour:2008gu}, 
\be
\label{eq:multipoles}
h_{\ell m}(x)=h_{\ell
  m}^{(N,\epsilon)}(x)\hat{S}^{(\epsilon)}(x)T_{\ell m}(x)e^{\ii
  \delta_{\ell m}(x)}(\rho_{\lm}(x))^\ell, 
\ee
where $h_{\ell m}^{(N,\epsilon)}(x)$ represents the Newtonian contribution
given by Eq.~(4) of~\cite{Damour:2008gu}, $\epsilon=0$ (or $1$) for 
$\ell+m$ even (odd). The remaining terms are defined as follows: 
$\hat{S}^{(\epsilon)}$ is the (specific) source, Eqs.~(15-16)
of~\cite{Damour:2008gu}; $T_{\ell m}$ is 
the tail factor that resums an infinite number of leading logarithms
due to tail effects, Eq.~(19) of~\cite{Damour:2008gu}; $\delta_{\ell m}$ 
is a residual phase correction, Eqs.~(20-28) of~\cite{Damour:2008gu};
and $\rho_{\ell m}$ is the residual amplitude
correction, that we keep up to 5PN fractional accuracy~\cite{Fujita:2010xj},
although their knowledge (and that of the $\delta_\lm$'s) has been recently
increased up to 14PN fractional order~\cite{Fujita:2011zk}. 

Note that the argument in the multipoles of Eq.~\eqref{eq:multipoles}
(and therefore in Eq.~\eqref{eq:rr}) is $x\equiv v_\varphi^2 = (r\Omega)^2$, 
that is preferable to $x_{\rm circ}\equiv \Omega^{2/3}$ 
due to the violation of the circular Kepler's constraint during 
the plunge phase~\cite{Damour:2006tr,Damour:2007xr}.
The sum in Eq.~\eqref{eq:Flux} is truncated at $\ell_{\rm max}=8$ included, 
and the system is initialized (in the strong-field region $6<r\leq 7$ )  
with post-circular initial data~\cite{Buonanno:2000ef,Nagar:2006xv}, 
which yields negligible initial eccentricity. 
The dynamics is then computed by solving Eqs.~(1)-(7) of Paper~I. 

\section{A hyperboloidal foliation of Schwarzschild spacetime}
\label{sec:layers}

In this Section we discuss the hyperboloidal layer approach
in Schwarzschild spacetime.
We construct a hyperboloidal foliation by gluing together a truncated Cauchy surface, 
which covers the strong-field region of the particle motion,  
and a hyperboloidal surface~\cite{Zenginoglu:2007jw}.  
Because a hyperboloidal surface is spacelike by construction, and
because Cauchy surfaces are also spacelike, one can choose a
global hyperboloidal foliation to agree with Cauchy surfaces in
a compact inner domain that includes the motion of the particle and
the central black hole.  
This choice allows us to employ standard coordinates near the central
black hole. The outer, asymptotic, domain is included in
the hyperboloidal layer.

\subsection{General properties}
A hyperboloidal layer is defined as a compact radial shell in which
the spacelike foliation approaches null infinity and the radial
coordinate is compactified. We determine the coordinates by requiring 
that outgoing null surfaces have the same representation 
in the layer coordinates as in the inner domain coordinates. 
We connect the coordinates used in the compact inner domain (Cauchy region) 
with the coordinates used in the outer domain (hyperboloidal layer) at an interface. 

\begin{figure}[t]
\center
\includegraphics[width=0.4\textwidth]{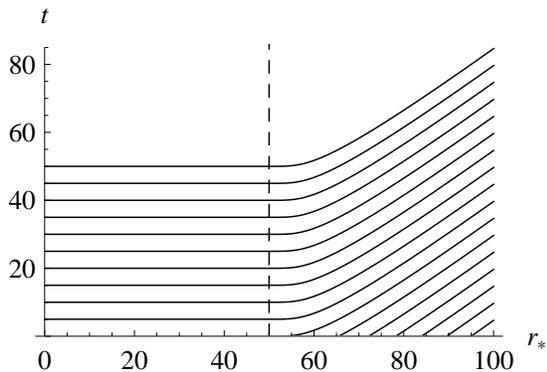}
\caption{Level sets of the hyperboloidal time $\tau$ as defined by
  Eqs.~\eqref{eq:tau_condition},~\eqref{eq:space} and~\eqref{compress}
  with respect to standard Schwarzschild coordinates $\{t,r_*\}$. The
  dashed line at $r_*=R_*=50$ depicts the location of the interface
  between the inner domain and the hyperboloidal layer.  
\label{trst}}
\end{figure}

We depict such a foliation with respect to standard coordinates
$\{t,r_\ast\}$ in Fig.~\ref{trst}. The level sets of the new time function,
$\tau(t,r_*)$, agree with the level sets of the standard
Schwarzschild time, $t$, for $r_\ast\leq R_*=50$. The dashed line
indicates the timelike surface, referred to as the interface
($r_\ast=R_*=50$), at which we smoothly modify the spacelike surfaces
to approach outgoing null rays asymptotically. 

\begin{figure}[t]
\center
\includegraphics[width=0.43\textwidth]{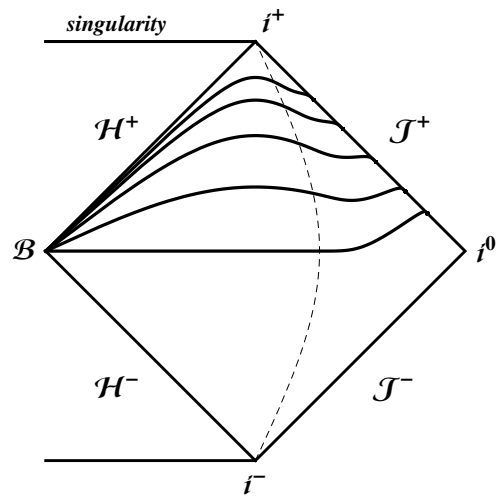}
\caption{Penrose diagram of Schwarzschild spacetime depicting the causal properties of the 
foliation plotted partially in Fig.~\ref{trst}. The dashed line indicates the
interface to the hyperboloidal layer. The time surfaces agree with
standard Schwarzschild time surfaces to the left of the interface. The
diagram also shows that the foliation stays spacelike everywhere,
including the asymptotic domain near null infinity.
\label{fig:penrose}}
\end{figure}

The spacelike surfaces partially depicted in Fig.~\ref{trst} approach outgoing null rays, 
but never become null surfaces themselves. The asymptotic causal
structure can not be clearly depicted in Fig.~\ref{trst}. A better
visualization of the causal structure is the Penrose diagram in
Fig.~\ref{fig:penrose}. The interface (still represented by a dashed
line) is depicted close to the black hole for visualization,
but the causal structure is accurate in this diagram. We see that the
hyperboloidal foliation agrees with standard $t$ surfaces near the
black hole. Beyond the interface, the surfaces smoothly approach
future null infinity in a spacelike manner. Although the surfaces look like they are
becoming null in Fig.~\ref{trst}, the Penrose diagram in
Fig.~\ref{fig:penrose} clearly shows that the surfaces are spacelike
everywhere. This causal behavior allows us to solve a usual 
initial-boundary value problem, while extracting
gravitational waveforms at future null infinity.

The hyperboloidal foliation that we employ is not only suitable for
wave extraction, it also provides a solution to the outer boundary
problem. Instead of truncating the simulation domain at a finite but
large distance, we employ a compactifying coordinate with respect to which 
null infinity is at a finite coordinate location. 
It is well known that compactification leads to loss of resolution
near the outer boundary when Cauchy foliations are
used~\cite{Orszag}. We do not run into this problem because we need to
resolve only a finite number of oscillations on an infinite domain
along hyperboloidal foliations, as opposed to an infinite number of
oscillations along Cauchy foliations~\cite{Zenginoglu:2010cq}. 

\begin{figure}[t]
\center
\includegraphics[width=0.33\textwidth]{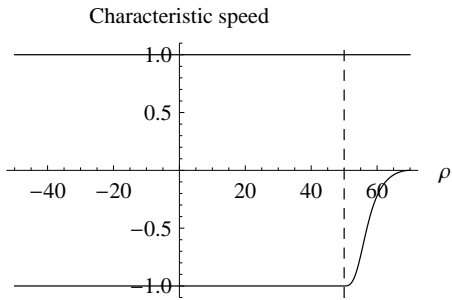}
\caption{Characteristic speeds on the numerical grid as given in Eq.~\eqref{speeds}. 
The dashed line denotes the location of 
the interface between the inner domain and the hyperboloidal layer. The outgoing speed 
has the same value in the inner domain as in the layer, whereas the incoming speed smoothy 
approaches zero in the layer.
\label{speeds}}
\end{figure}

A good illustration that compactification solves the outer boundary
problem is given by a depiction of characteristic speeds on the
numerical grid (Fig.~\ref{speeds}). The outgoing speed of
characteristics is nonvanishing finite at future null infinity. The
incoming speed, on the other hand, vanishes because future null
infinity is itself an incoming null surface
(Fig.~\ref{fig:penrose}). No outer boundary conditions are needed 
because there are no incoming characteristics from the outer boundary.   

Note that the compactifying coordinate is conceptually independent
from the hyperboloidal foliation. We can choose any compactifying
coordinate along the spacelike surfaces of our foliation compatible
with scri-fixing. The choice of hyperboloidal foliation and
compactification together determines the structure of characteristics
on the numerical grid. The choices for Fig.~\ref{speeds} ensure that
the outgoing characteristic speed is unity in the layer. In the next
Section we discuss how to achieve this. 

\subsection{Explicit construction of the hyperboloidal layer} 
\label{sec:layer}

There are different ways to construct a hyperboloidal layer. 
One we find most lucid is to consider the expression of outgoing null 
rays in local coordinates. In standard waveform extraction methods, 
the solution is computed along $t$ surfaces and the waveform is plotted along 
the outgoing null surfaces $t-r_\ast$. 
Naturally, we would like to keep the expression of outgoing null rays
invariant in our formulation. We would also like to keep the time
direction invariant, so that ringdown frequencies or decay rates that
we compute are physical. Our requirements for a suitable  
hyperboloidal layer are as follows:
\begin{enumerate}
\item The exterior timelike Killing vector field
  in local coordinates is kept invariant in the layer. 
\item The outgoing null rays in local coordinates is
  kept invariant in the layer. 
\item The local coordinates in the layer agree with the standard
  $\{t,r_\ast\}$ coordinates at the interface. 
\end{enumerate}

Now we formalize these requirements. The first requirement gives a
relation between the new time coordinate $\tau$ and the standard time
coordinate $t$. The requirement that the Killing field is kept
invariant translates into $\partial_t=\partial_\tau$. This condition
is fulfilled by a transformation of the form
\be
\label{eq:tau_transf}
\tau = t-h(r_*) \ ,  
\ee
where the function $h(r_*)$ is called the \emph{height function}. The
height function can only depend on spatial coordinates to leave the
timelike Killing field invariant. We let the height function depend
only on the tortoise coordinate because our problem is spherically
symmetric. 

Under the transformation~\eqref{eq:tau_transf} the Schwarzschild 
metric~\eqref{eq:ss} becomes
\be
\label{eq:hypmetric} 
g = A\left(-d\tau^2 - 2H d\tau dr_\ast
+\left(1-H^2\right) dr_\ast^2\right) + r^2 d\sigma^2,
\ee
where $H\equiv dh/dr_{\ast}$ is called the \emph{boost function}. For
example, ingoing Eddington-Finkelstein coordinates are obtained with $H = -2/r$.
Similarly, Painlev\'e-Gullstrand coordinates are obtained 
with $H = -\sqrt{2/r}$. The constant time hypersurfaces in these coordinates 
foliate the event horizon instead  of intersecting at the bifurcation sphere 
and are therefore suitable for excision. Note that both choices give 
$H=-1$ at the horizon~\cite{Zenginoglu:2011jz}.

We require an analogous behavior in the asymptotic domain, in the sense that 
the resulting surfaces should foliate future null infinity instead of intersecting 
at spatial infinity. The analogy with excision indicates that one needs to satisfy 
$H=1$ at infinity. 
The choice of a suitable boost function follows from the second item in our list.
We require that the outgoing null rays in local coordinates is kept invariant. 
Denoting the layer coordinates with $\{\tau, \rho\}$, we require
\be\label{eq:tau_condition}
t - r_*= \tau - \rho\,,
\ee
where $\rho$ is a yet unspecified compactifying coordinate. By combining 
Eqs.~\eqref{eq:tau_transf} and~\eqref{eq:tau_condition} we get for the 
height function $h(r_*)=r_*-\rho(r_*)$. Taking the derivative of this equation with 
respect to $r_*$, we obtain the following relation between the boost function
$H$ and the Jacobian $d \rho(r_\ast)/dr_*$ of the spatial compactification
\be
\label{eq:relation} 
\dfrac{d\rho}{dr_*} = 1-H\,.
\ee
The Jacobian of any compactification vanishes at the domain boundary, 
so we have $H=1$ at null infinity.

The condition~\eqref{eq:tau_condition} has two important consequences. 
First, the outgoing characteristic speed, which is $+1$ in the inner domain, remains 
$+1$ also across the hyperboloidal layer. Second, the incoming characteristic speed, 
which is $-1$ in the inner domain, smoothly decreases in the layer to reach zero at 
future null infinity. This is easily seen by writing the Schwarzschild
metric~\eqref{eq:hypmetric} using the compactifying coordinate $\rho$  
as\footnote{Note that the metric in Eq.~\eqref{eq:compmetric} is
  singular at the boundary because the Jacobian of any
  compactification is singular. This singularity can be rescaled away
  with a conformal factor, but such a rescaling is not necessary for
  our purposes because the RWZ equation in hyperboloidal
  compactification is regular without an explicit conformal
  rescaling of the background~\cite{Zenginoglu:2009ey}.}  
\be
\label{eq:compmetric} 
g = A\left(-d\tau^2 - \dfrac{2H}{1-H} d\tau d\rho + \dfrac{1+H}{1-H} 
d\rho^2\right) + r(\rho)^2 d\sigma^2.
\ee
The outgoing ($c_+$) and incoming ($c_-$) characteristic speeds of 
spherically symmetric null surfaces read 
\be
\label{eq:speeds} c_+ =  1, 
\qquad c_-= - \frac{1-H}{1+H}. 
\ee
Note that $c_-=0$ at the outer boundary of the $\rho$-domain, 
where $H=1$ by \eqref{eq:relation}. 
The speeds are plotted in Fig.~\ref{speeds} 
for a particular choice of spatial compactification that we 
describe in Sec.~\ref{sec:comp}. 

\begin{figure}[t]
\center
\includegraphics[width=0.4\textwidth]{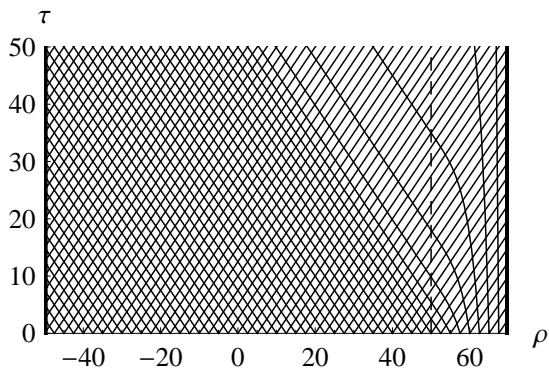}
\caption{The structure of the characteristics on the numerical grid. 
         Compare Figs.~\ref{trst}, \ref{fig:penrose}, and \ref{speeds}.
\label{chars}}
\end{figure}

Let us now discuss the third condition in our list, namely the requirement that the 
new coordinates $\{\tau,\rho\}$ agree with standard coordinates $\{t,r_*\}$ at 
the interface between the inner domain and the hyperboloidal layer. This condition
can be fulfilled by a suitable choice of the compactifying coordinate. 
The spatial compactification $r_\ast(\rho)$ shall have the 
following differentiability properties along the interface at $r_\ast = \rcm$
\begin{align}
\label{eq:cond1}
r_\ast(\rcm) &= \rcm,                                              \\
\label{eq:cond2}
\left.\frac{d r_\ast}{d\rho}\right\vert_{\rho=\rcm} &= 1,             \\
\label{layer}
\left.\frac{d^kr_\ast}{d\rho^k}\right\vert_{\rho=\rcm} &=0\,, \quad k>1.  
\end{align}
These relations imply that the coordinates $r_\ast$ and $\rho$, and therefore $t$ and $\tau$, 
agree along the interface to $k$th order. We give an explicit choice for the compactification 
$r_*(\rho)$ in Sec.~\ref{sec:comp}.

For completeness, we finally depict in Fig.~\ref{chars} the global
structure of the characteristics propagating along the numerical grid
$\{\tau,\rho\}$ in a suitable hyperboloidal compactification. The
outgoing characteristics are straight lines with 45 degrees to the
$\rho$-axis, just as for the $\{t,r_\ast\}$ coordinates. In agreement
with Eq.~\eqref{eq:speeds} and Fig.~\ref{speeds}, there are no
incoming characteristics from the outer boundary.  

A point where our approach can be further improved is indicated in Fig.~\ref{chars}.
We truncate the infinite computational domain in $r_*$ to the left 
arbitrarily at $r_*=-50$. As a result, there are incoming modes from
the inner boundary that need to be set by artificial boundary
conditions.  This procedure can contaminate the interior solution and make the calculation of 
GWs absorbed by the black hole inaccurate (for example, to reduce contamination 
Ref.~\cite{Martel:2003jj} uses a very large value of the extraction
radius, $r_*^{\rm{extr}}=-1500M$, for computing the absorbed fluxes). 
In addition, the efficiency of the numerical computation is reduced by the 
coordinates used near the black hole.
We accept these disadvantages because we want to describe the dynamics of 
the test-mass using $\{t,r_*\}$ coordinates, exactly as in Paper I. 

A way to avoid the inner timelike boundary
near the black-hole horizon is to work in horizon-penetrating coordinates
in combination with excision. Then one needs to transform the 
RWZ equations, their sources, as well as the relative dynamics of the binary, 
that we used in Paper I, to horizon-penetrating coordinates 
(coordinate-independent expressions for the RWZ equations and sources 
are explicitly given in Ref.~\cite{Sarbach:2001qq,Martel:2005ir}). 
Using a horizon-penetrating, hyperboloidal foliation 
is the cleanest option to compute accurately both the asymptotic and the absorbed waves. 
Alternatively, one can construct such coordinates also
by attaching an internal layer to the truncated $\{t,r_*\}$ domain 
so that the event horizon, $r_*=-\infty$,  is compactified. Because 
our main focus in this study is on the asymptotic waveform, we use 
the hyperboloidal layer only in the exterior asymptotic domain.

\subsection{Spatial compactification} 
\label{sec:comp}

We present the form of the compactifying coordinate that we use in our 
numerical calculations. We transform $r_*$ by introducing a compactifying 
coordinate $\rho$ via
\be
\label{eq:space} r_\ast = \frac{\rho}{\Omega(\rho)},
\ee
where, $\Omega(\rho)$ is a suitable function of $\rho$ (not to be confused with 
the orbital frequency in Sec.~\ref{sec:rd}). 
The function $\Omega(\rho)$ has similar properties 
as the conformal factor in the conformal compactification of asymptotically 
flat spacetimes  proposed by Penrose \cite{Penrose63, Penrose65}. 
For the regularity of the transformation in the interior we require that 
$\Omega$ has a definite sign, say, $\Omega>0$ for all $\rho<S$, 
where $S$ denotes the coordinate location of null infinity, and therefore 
the zero set of $\Omega$. To map the infinite domain $R_*\leq r_*<+\infty$ 
to the finite domain $R_*\leq  \rho \leq S$ we require
\be\label{penrose}  \Omega(S) = 0, \quad \Omega'(S) \ne 0.  \ee
where $\Omega'\equiv d\Omega/d\rho$.

In addition, we also require that our coordinates agree with standard 
coordinates in an inner domain. Therefore we set $\Omega =1$ for all $\rho\leq R_*$, 
where $R_*$ denotes the location of the interface. The transition to the 
layer at this interface needs to be sufficiently smooth for a stable 
numerical implementation. We require 
in accordance with Eqs.~\eqref{eq:cond2}-\eqref{layer}
\be
\frac{d^k\Omega}{d\rho^k}\Big|_{\rho=\rcm} = 0 \quad{\rm with}\quad k\geq 1. 
\ee
The maximum value of $k$ for which the above property 
is satisfied determines the differentiability of the layer.

By differentiating Eq.~\eqref{eq:space}, we get with Eq.~\eqref{eq:relation}
\be 
\label{eq:boost} 
H(\rho)=1 - \dfrac{\Omega^2}{\Omega-\rho\, \Omega'}\,. 
\ee

The form of the compactifying coordinate~\eqref{eq:space} is convenient 
because it allows us to control the hyperboloidal foliation 
by a suitable function $\Omega(\rho)$ via Eq.~\eqref{eq:boost}.
It also makes the connection to the definition of asymptotic flatness
within the Penrose conformal compactification picture clear. 
However, we emphasize that, in our specific case, we can also use a more
general transformation than~\eqref{eq:space}, which fulfills the conditions of a 
coordinate compactification.

\section{The RWZ equations}
\label{sec:RWZ}

In this Section we discuss the RWZ equations as implemented numerically.
For the relations of the RWZ master function with the asymptotic 
observable quantities see Appendix~\ref{app:asymp}.

\subsection{The RWZ equations in the interior}
\label{sbsc:RWZ_cauchy}

In the interior domain the RWZ equations with a
point-particle source are written as in Paper I. Given the dynamics of
the particle, one solves the following two decoupled partial
differential equations for each multipole  $(\ell,m)$ of even (e) or
odd (o) type\footnote{In our case, these correspond respectively 
to multipoles with $\ell+m={\rm even}$ and $\ell+m={\rm odd}$.} 
\begin{equation}
\label{eq:rwz}
\de_t^2\Psi^{(\rm e/o)}_{\ell m}-\de_{r_*}^2\Psi^{(\rm e/o)}_{\ell m} + V^{(\rm
  e/o)}_{\ell}\Psi^{(\rm e/o)}_{\ell m} = {\cal S}^{(\rm e/o)}_{\ell m}, 
\end{equation}
with source terms ${\cal S}^{(\rm e/o)}_{\ell m}$ that are explicit functions
of the phase-space variables $(r_*,p_*)$. 
The sources have the structure 
\begin{align}
\label{source:standard}
{\cal S}^{(\rm e/o)}_{\ell m} & = G^{(\rm e/o)}_{\ell m}(r,t)\delta(r_*-r_*(t)) \nonumber \\
& + F^{(\rm e/o)}_{\ell m}(r,t)\de_{r_*}\delta(r_*-r_*(t)) \ ,
\end{align} 
where $r_*(t)$ is here indicating the particle radial coordinate.
The explicit expressions for the sources are given in 
Eqs.~(20)-(21) of~\cite{Nagar:2006xv}, to which we address 
the reader for further technical details. In our approach the distributional 
$\delta$-function is approximated by a narrow Gaussian of 
finite width $\sigma\ll M$ (see Sec.~\ref{sec:particle}).

\subsection{The RWZ equations in the hyperboloidal layer}
\label{sbsc:RWZ_layers}

As explained in Sec.~\ref{sec:layers} there are three essential steps to the
construction of the hyperboloidal layer: 
\begin{enumerate}
\item Introduce a new time coordinate $\tau$, Eq.~\eqref{eq:tau_transf}, that preserves the 
stationarity of the background,
\be
\partial_t = \partial_\tau \quad \Rightarrow \quad \tau = t - h\,.
\ee
\item Fix the time coordinate such that the expression of the outgoing
  null rays is invariant in the layer, 
\be
\label{eq:fundamental}
t-r_\ast = \tau - \rho \quad \Rightarrow \quad H = 1 - \dfrac{d\rho}{d r_*}\,.
\ee
\item Choose a suitable compactifying coordinate $\rho$ so that the coordinates 
in the layer agree with the coordinates near the black hole,
satisfying the conditions~\eqref{eq:cond1}-\eqref{layer}.
\end{enumerate}
The whole prescription results in is a simple coordinate
transformation, $\{t,r_\ast\}\to\{\tau,\rho\}$, 
that satisfies the above properties. 
The derivative operators in standard coordinates transform as
\be
\partial_t = \partial_\tau, \qquad \partial_{r_\ast} = -H\,\partial_\tau + (1-H) \,\partial_\rho\,. 
\ee
Applying this transformation on Eq.~\eqref{eq:rwz}  (dropping all multipolar indices)
\be \label{eq:RWZ_dropped}
(\de_t^2-\de_{r_*}^2 + V) \Psi = \S, 
\ee
we get for the wave operator in the new coordinates
\begin{eqnarray*} 
&&\partial_t^2 - \partial_{r_*}^2 = -(1-H^2)\partial_\tau^2 + \\ 
&&+(1-H)\left( -2H\partial_\tau \partial_\rho +
  (1-H)\partial_\rho^2-(\partial_\rho H) (\partial_\tau +
  \partial_\rho) \right). 
\end{eqnarray*}
We can take out a $(1-H)$ term from the operator. We need to be
careful with the lower order terms in \eqref{eq:RWZ_dropped}. The
source term is compactly supported in a neighborhood of the particle
in the interior domain and therefore is not a concern. The potential,
however, is nonvanishing in the wave zone. Its fall-off behavior is
essential for the applicability of the hyperboloidal
method~\cite{Zenginoglu:2009ey}.  
The potential in the RWZ equation falls off as $r^{-2}$ both for even
and odd parity perturbations.  
Therefore we can introduce the rescaled potential
\be
\label{resc_pot} 
\bar{V}\equiv V/(1-H), 
\ee
which has a regular limit at null infinity. To see this, consider for example
the odd-parity (Regge-Wheeler) potential
\be 
V^\textrm{(o)}= \frac{1}{r^2}\left(\ell(\ell+1)-\frac{6}{r}\right)\ ,
\ee
we have with \eqref{eq:boost}
\be 
\bar{V}^{\rm (o)} =   \frac{V^{\rm (o)}}{1-H} = 
\frac{(\Omega - \rho\, \Omega')}{\rho_r^2}\left(\ell(\ell+1)-\frac{6 \Omega}{\rho_r}\right) \ ,
\ee
where $\rho_r\equiv\Omega\, r$. The rescaled Schwarzschild radius $\rho_r$ has a nonvanishing
limit at infinity because $r$ and $r_\ast$ coincide asymptotically. 
As a result, we have  $\rho_r = \rho=S$ at infinity. 
An analogue regular expression holds also for the even-parity (Zerilli) potential.

Then we can write the RWZ equation in the layer as
\begin{eqnarray}\label{eq:rwz_layer} 
&& -(1+H)\partial_\tau^2 \Psi -2H\partial_\tau \partial_\rho \Psi +
  (1-H)\partial_\rho^2 \Psi \nonumber \\ 
&& -(\partial_\rho H) (\partial_\tau + \partial_\rho)\Psi + \bar{V}\Psi = 0.
\end{eqnarray}
From this form of the equation, it is immediately clear that setting $H=0$ recovers 
the standard RWZ equation~\eqref{eq:rwz}. We also see that the equation is regular 
and pure outflow at infinity ($H=1$).

\section{Numerics}

The numerical technique employed in our code is a standard combination
of finite-difference approximation for the spatial derivatives and
Runge-Kutta methods for time integration~\cite{Zenginoglu:2009ey,Bernuzzi:2010ty}.  
In this section we briefly review the method.

\subsection{Numerical methods} 

Our code solves the RWZ equation in first-order-in-time
second-order-in-space form adopting the method of lines and the Runge-Kutta
4th order scheme. The right hand side is discretized in space on a uniform grid in
the coordinate $\rho\in[\rho_{\rm min},S]_{\rcm}$, where
$\rcm$ denotes the interface to the hyperboloidal layer
and $S$ the coordinate location of $\scri^+$. 
Finite differences are employed for the derivatives. We use
4th order central stencils in the bulk, lop-sided or sided 4th order
stencils for the outermost points ($\rho=\rho_{\rm min}$ and $\rho=S$).
No boundary data is prescribed at $\scri^+$, whereas maximally
dissipative 4th order convergent outgoing boundary 
conditions~\cite{Calabrese:2005fp} are imposed at the inner boundary.
Kreiss-Oliger type dissipation is added to the RWZ equation. 
The particle trajectory is updated using a 4th order Runge-Kutta integrator 
with adaptive time-step. The convergence of the code is demonstrated in
Appendix~\ref{app:conv}. 

In our numerical computations we set
\be\label{compress} \Omega = 1- \left(\frac{\rho-\rcm}{S-\rcm}\right)^4
\Theta(\rho-\rcm)\,,\ee
though various other choices are possible.
The step function, $\Theta(\rho-\rcm)$, indicates that compactification
is performed only for $\rho>\rcm$. We choose the numerical domain as
$[\rho_{\min},S]_{\rcm}=[-50,70]_{50}$; Figs.~\ref{trst}, \ref{speeds},
and~\ref{chars} refer to these settings.
For the production runs that we present below, the $\rho$-domain is covered by 
12001 points, that correspond to gridspacing $\Delta\rho=0.01$.

\subsection{Particle treatment}
\label{sec:particle}

Following previous work~\cite{Nagar:2006xv,Bernuzzi:2010ty}, 
the $\delta$-function in the RWZ source is represented  by 
a narrow Gaussian of finite width $\Delta \rho<\sigma\ll M$.
The hyperboloidal compactification has an advantage also on the
treatment of the Dirac distribution via a smooth Gaussian 
because most of the computational resources are used for the 
strong-field, bulk region so that narrow Gaussians can be 
efficiently resolved. For the production runs that we present 
below, we use $\sigma=0.08M$.

We inject zero initial data for the RWZ master functions 
switching on the sources progressively in time~\footnote{This approach has been suggested 
to reduce the impact of 
Jost solutions~\cite{Field:2010xn,Canizares:2010yx,Jaramillo:2011gu}.} 
following the prescription~\cite{Field:2010xn},
\be
S \mapsto \frac{S}{\exp\left[-a_0(t-t_0)\right]+1 } \ ,
\ee
where typically $a_0=1/\,M$ and $t_0=40\,M$. We observed that this smooth 
switch-on significantly reduces the  (localized)``junk'' radiation contained 
in the initial data, without, obviously, eliminating it completely.

\section{Results}

Let us briefly summarize our main results.
In Sec.~\ref{sbsec:ciro} we focus on circular orbits 
to assess the performance of our new numerical implementation.
We compute the gravitational energy flux emitted at null infinity 
by a particle on stable circular orbits and compare it with the 
semi-analytic data of Fujita et al.~\cite{Fujita:2004rb}. 
We also compute (and characterize) the GW energy flux emitted 
by the particle on unstable circular orbits. In particular, 
we extract from the data the corresponding residual amplitude 
corrections $\rho_\lm$ introduced in Ref.~\cite{Damour:2008gu}.
We focus then on the transition from quasi-circular inspiral to plunge,
merger and ringdown. In Sec.~\ref{sbsec:inspl} we discuss the total gravitational 
waveform, including up to $\ell_{\max}=8$ multipoles, extracted at $\scri^+$.
This waveform is then compared in Sec.~\ref{sbsc:finitextr} to waveforms 
extracted at finite radii. We estimate phase and amplitude differences 
and test the standard extrapolation procedure that is routinely 
applied to NR waveforms.
In Sec.~\ref{sbsec:angmom} a self-consistency check of the treatment 
of the dynamics is presented. Our prescription for the 
radiation reaction is checked on consistency (even beyond the LSO crossing) 
between the GW angular momentum flux extracted at $\scri^+$  and the 
(5PN EOB-resummed) mechanical angular momentum loss $\F_\varphi$.
In Sec.~\ref{sbsc:kick} we compute the final and maximum gravitational recoil 
of the final black-hole in the $\nu\to 0$ limit, obtaining a more accurate 
estimate than the ones given in Paper I.

\subsection{Circular orbits}
\label{sbsec:ciro}

\subsubsection{Accuracy: comparison with data by Fujita et al.}

\begin{figure}[t]
\center
\includegraphics[width=0.46\textwidth]{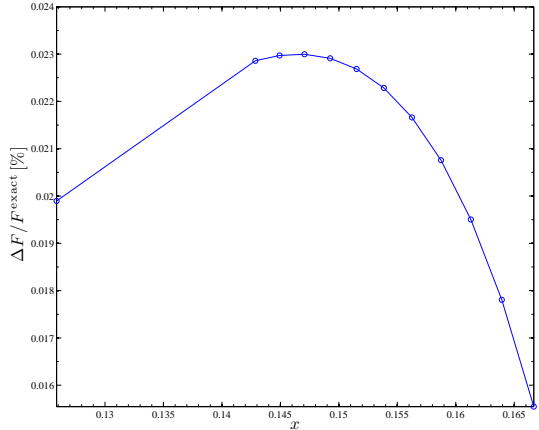}
\caption{\label{fig:flux_comp_fuji}Stable circular orbits: fractional
  difference between the RWZ total energy flux computed with our code
  and extracted at $\scri^+$ (up to $\ell=8$) and the corresponding 
  semi-analytic data computed by Fujita et al.~\cite{Fujita:2004rb}.}
\end{figure}

As a test of the accuracy of our new setup we compute
the gravitational wave energy and angular momentum fluxes 
emitted by a particle on stable circular orbits. 
For each orbital radius, $r_0$ (in units of $M$ hereafter), we consider 
the complete multipolar waveform (up to $\ell_{\rm max}=8$) measured at $\scri^+$ 
and compute the fluxes summing together all multipoles via Eqs.~\eqref{eq:dEdt} and~\eqref{eq:dJdt}.  
We consider circular orbits belonging to both the stable branch 
($r_0\geq 6$) and the unstable branch ($3<r_0<6$). 
The computation of the GW fluxes from stable 
circular orbits in Schwarzschild spacetime has been performed several times 
in the past, with different integration techniques (either in time domain or in frequency domain) 
and with increasing level of accuracy~\cite{Cutler:1993vq,Pons:2001xs,Martel:2003jj,
Fujita:2004rb,Sopuerta:2005gz,Yunes:2008tw}.
Currently, the method that yields the most accurate results is the one developed 
by Fujita et al.~\cite{Fujita:2004rb}, which allows for the computation of 
emitted fluxes with a relative error of order $10^{-14}$.
We checked the accuracy of our numerical setup 
(finite differencing with a hyperboloidal layer and wave extraction at $\scri^+$) 
by considering  a small sample of stable orbits, with radii 
in the range $6\leq r_0 \leq 7.9456$ and spaced by $\Delta r_0=0.1$ 
for $6\leq r_0 \leq 7$. The full multipolar information for $r_0=7.9456$ 
(both energy and angular momentum fluxes) is listed in Table~\ref{tab:gwflux} 
in Appendix~\ref{app:conv}, so to facilitate the comparison  
with published data~\cite{Martel:2003jj,Sopuerta:2005gz}.
In addition, a direct comparison with the data kindly given to us by Ryuichi Fujita 
and computed as in Ref.~\cite{Fujita:2004rb}, that we consider ``exact'', 
reveals that our finite-differencing, time-domain computation 
is rather accurate: The relative difference 
$\Delta F_{\ell m}/F_{\ell m}^{\rm Exact}=(F_\lm^{\rm RWZ}-F_\lm^{\rm Exact})/F_\lm^{\rm Exact}$ 
in energy flux  is below $0.8$~\% in almost every multipolar channel
(see Appendix~\ref{app:conv} for more detailed information).
Summing together all multipoles, we find that the total energy flux, 
dominated by the modes with smaller values of $\ell$ and with $m=\ell$,  
agrees with the exact data within $0.02$~\%. In Fig.~\ref{fig:flux_comp_fuji} 
we show the relative difference between total fluxes, 
$\Delta F/F^{\rm Exact}=(F^{\rm RWZ}-F^{\rm Exact})/F^{\rm Exact}$ 
(summed up to $\ell_{\rm max}=8$),
versus $x=1/r_0$. 

\subsubsection{Total energy flux, unstable orbits and the ``exact''  
               multipolar amplitudes $\rho_{\ell m}$}

\begin{figure}[t]
\center
\includegraphics[width=0.46\textwidth]{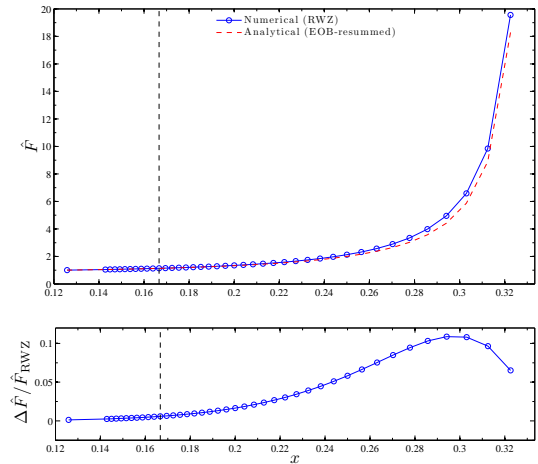}
\caption{\label{fig:hatf_ciro} Newton-normalized total gravitational wave energy 
  flux summed up to $\ell=8$. The analytical (5PN-accurate, EOB-resummed) flux 
  is compared with the numerical points, that include also unstable circular orbits. 
  The vertical dashed line indicates the LSO location at $x=1/6$.}
\end{figure}

\begin{figure*}[t]
\center
\includegraphics[width=0.46\textwidth]{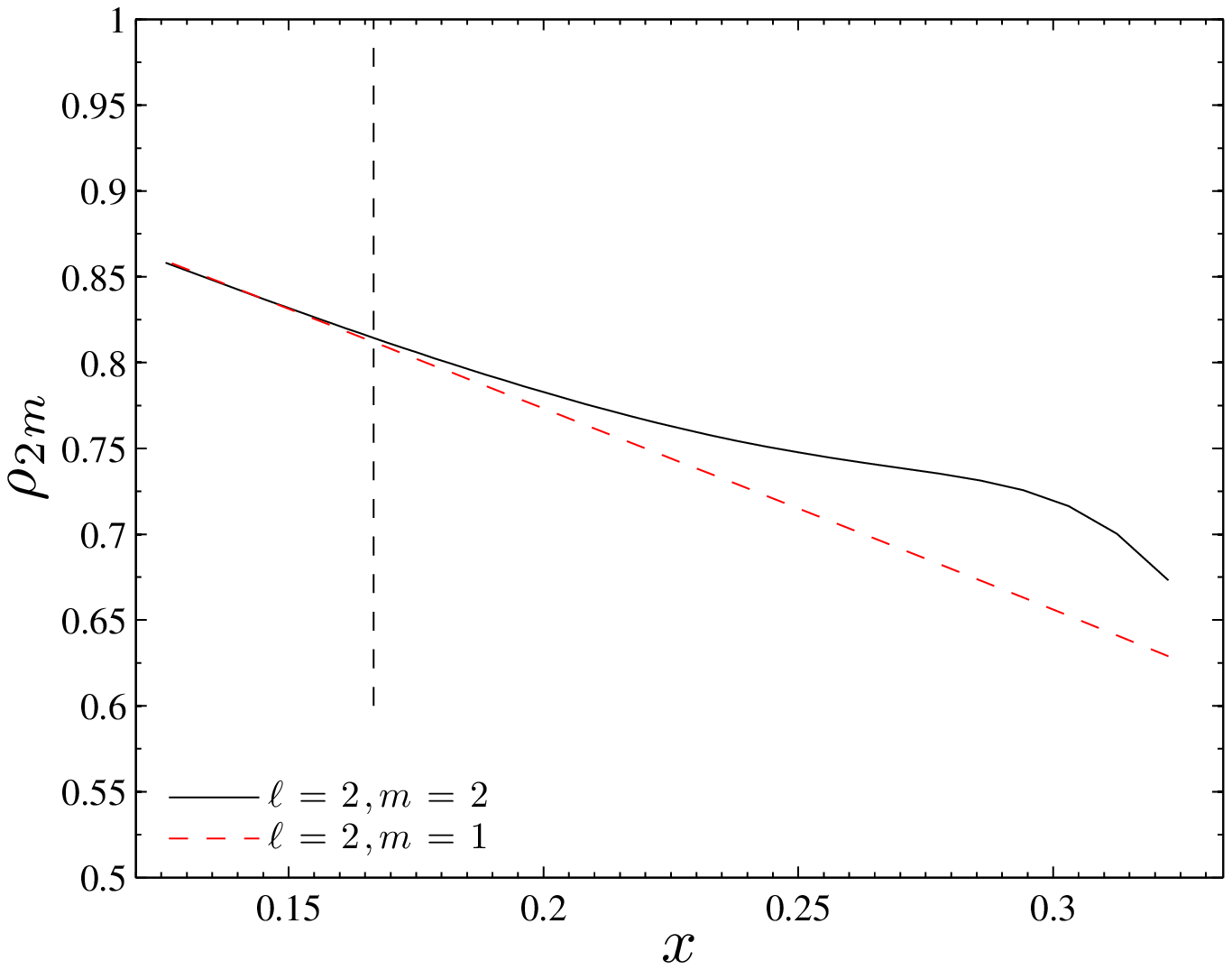}
\includegraphics[width=0.46\textwidth]{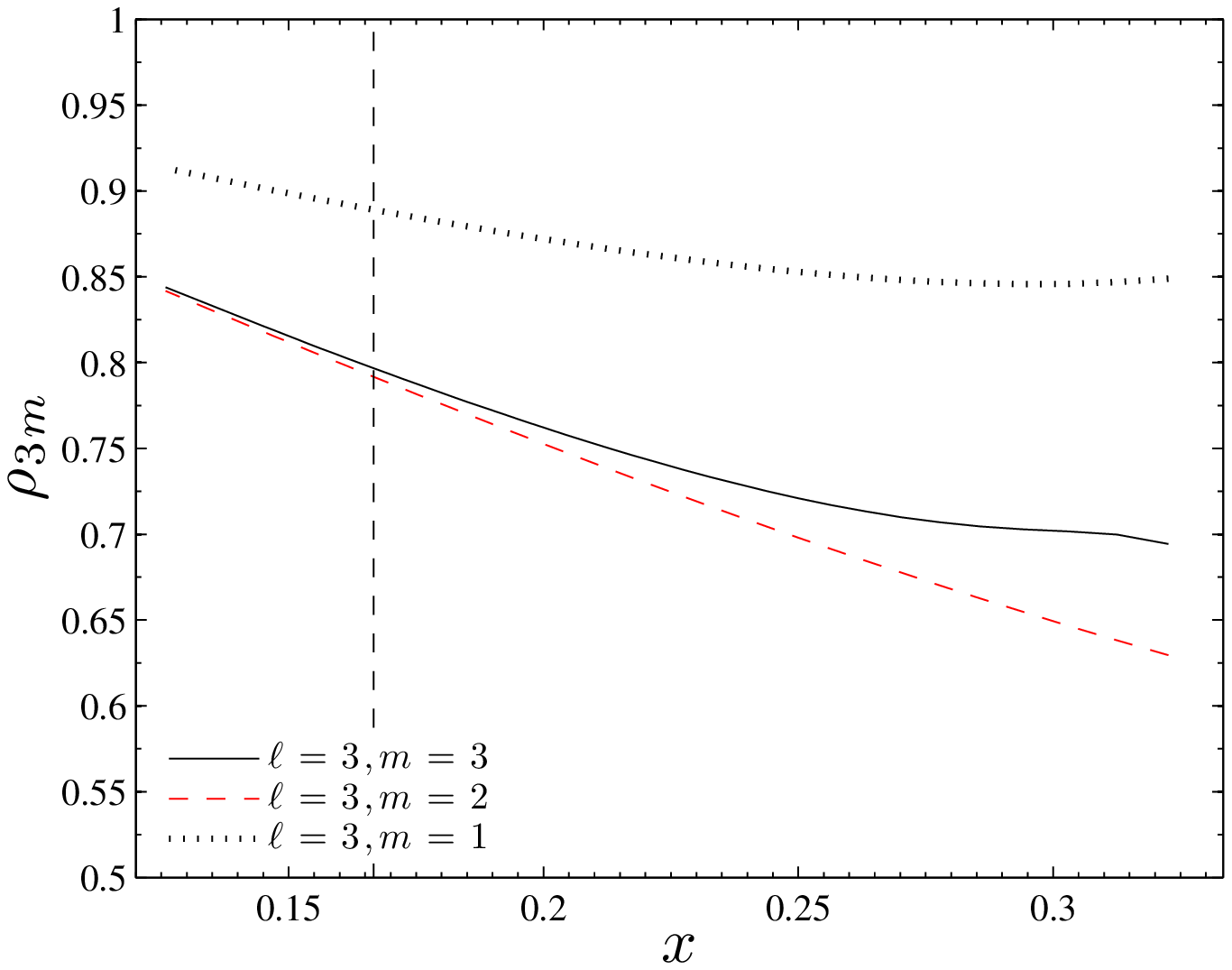}\\
\includegraphics[width=0.46\textwidth]{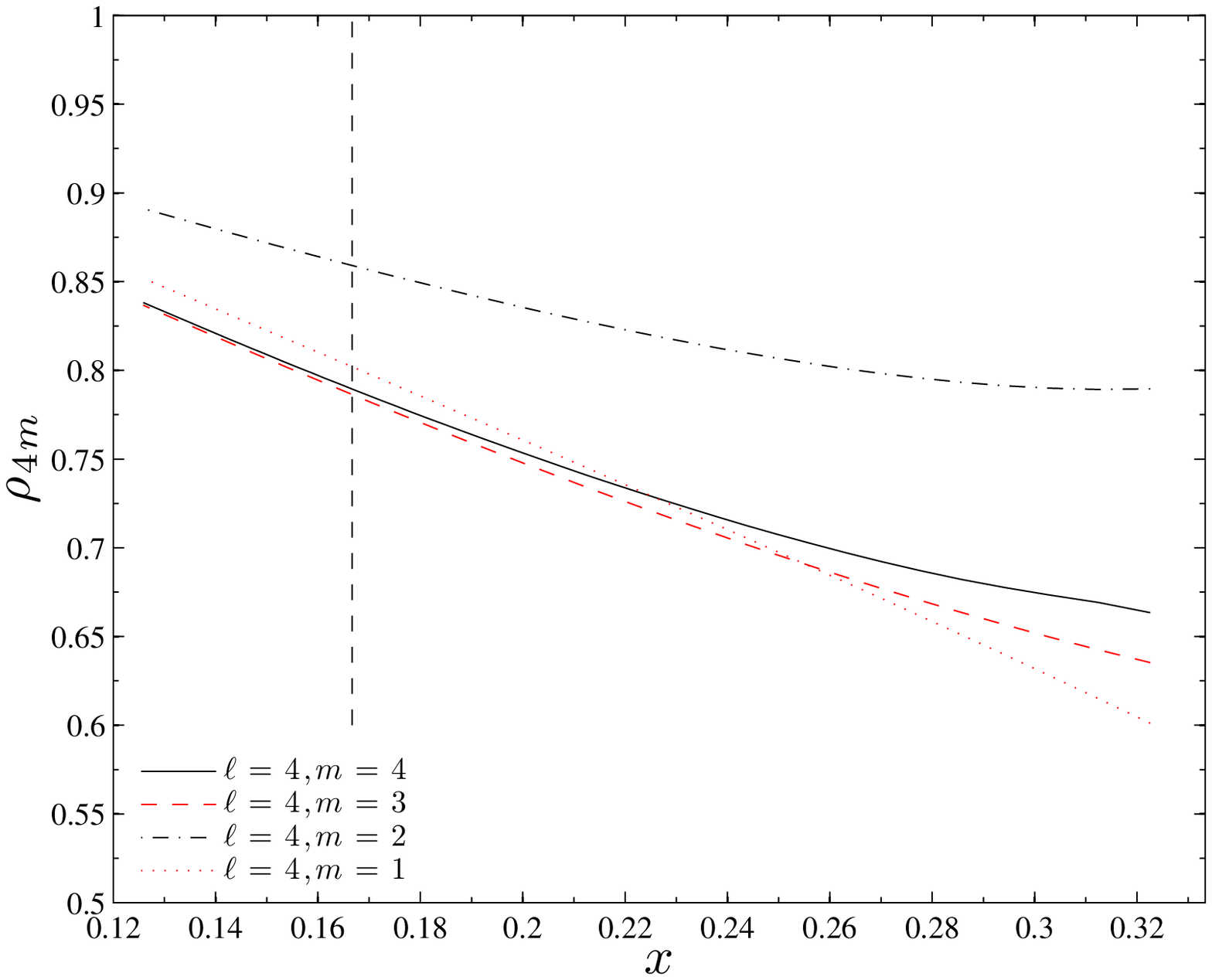}
\includegraphics[width=0.46\textwidth]{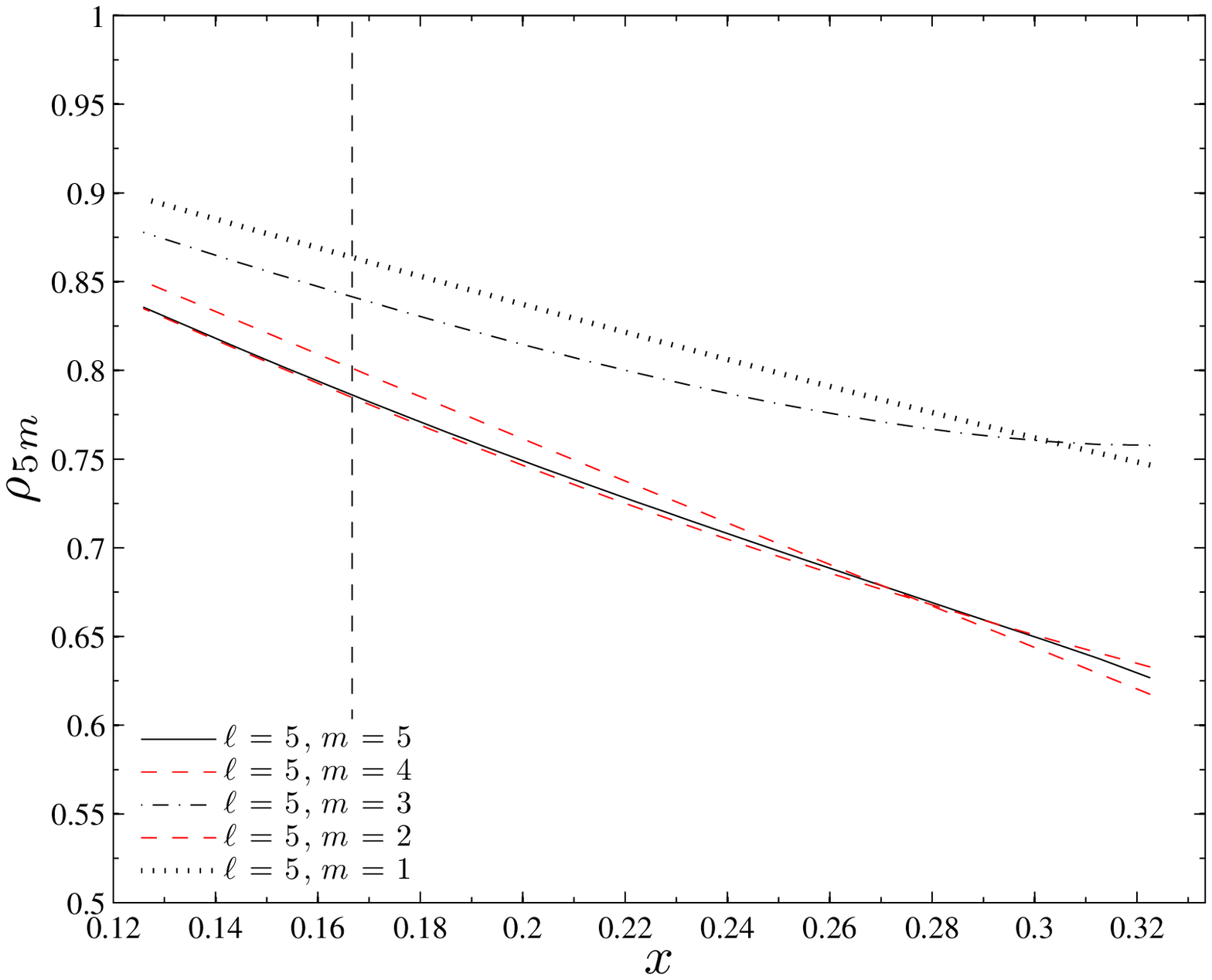}\\
\includegraphics[width=0.46\textwidth]{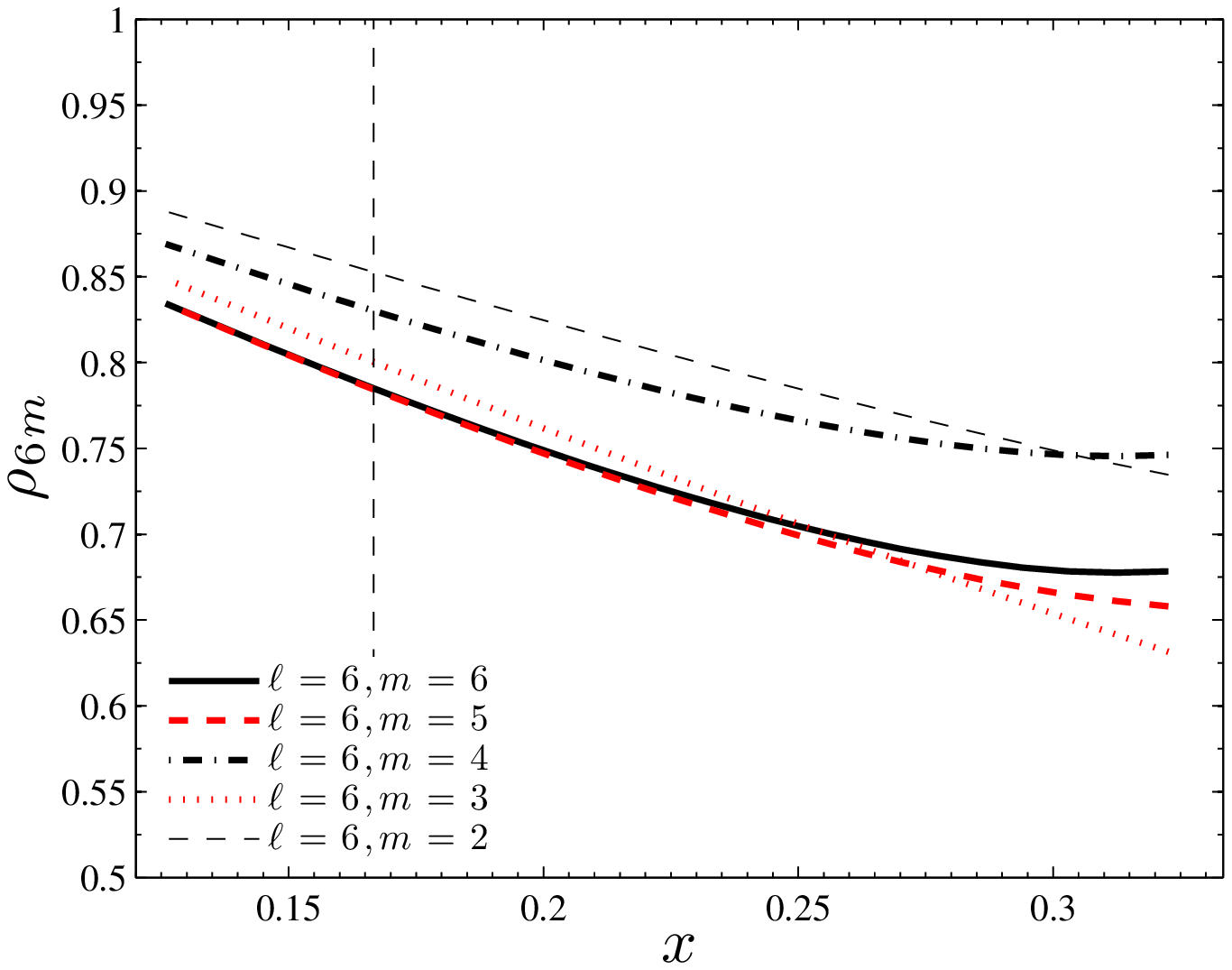}
\includegraphics[width=0.46\textwidth]{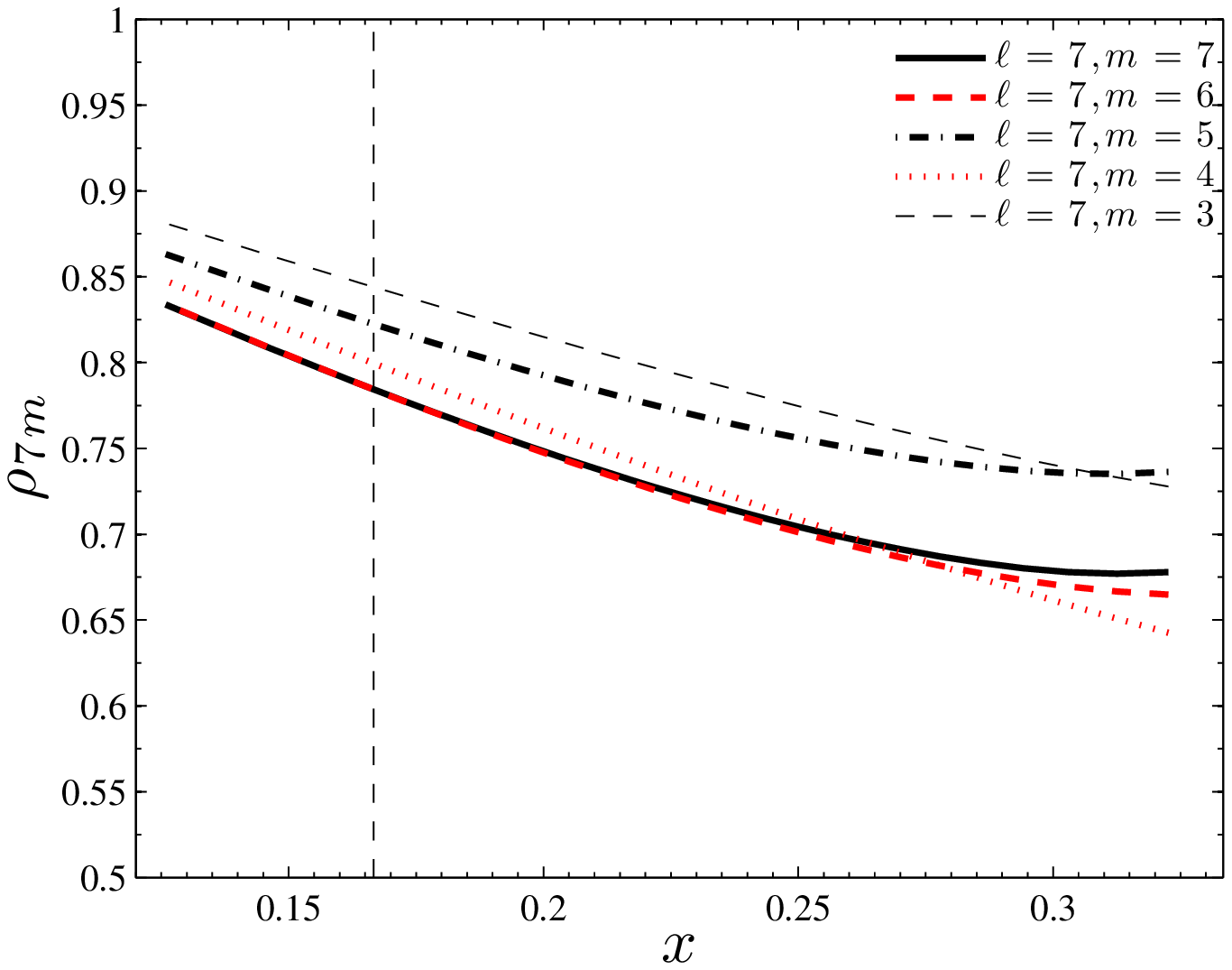}\\
\caption{\label{fig:rholm_ciro}The ``exact'' functions $\rho_{\lm}$ extracted 
from the numerical fluxes for $1/7.9456\leq x \leq 1/3.1 $. The vertical dashed
line indicates the LSO location, $x=1/6$.}
\end{figure*}

Now that we have assessed the accuracy of our finite-difference, time-domain 
code, we calculate the GW energy flux for {\it unstable} circular orbits, 
i.e. orbits with radii in the range $3<r_0<6$. 
This computation has received a rather poor attention in the literature. 
To our knowledge, the only computation along unstable orbits was performed 
in Ref.~\cite{Damour:2008gu} for the $\ell=m=2$ flux 
and with less good accuracy than what we are able to do here.
In~\cite{Damour:2008gu} it was pointed out that the knowledge of the emitted flux also
{\it below} the LSO might be helpful to improve the resummation of the residual amplitude
corrections $\rho_{\ell m}$ that enter the factorized (EOB-resummed) multipolar waveform
introduced there.

We compute the multipolar fluxes for a sample of unstable circular orbits with 
$3.1\leq r_0 <6$, spaced by $\Delta r_0=0.1$.
Figure~\ref{fig:hatf_ciro} shows in the top panel (as a solid line with circles) the 
two branches together, both for stable and unstable orbits, of the Newton-normalized 
total energy flux, $\hat{F} = F_\lm/F^N_{22}$, summed over all
multipoles up to $\ell_{\rm max}=8$.
The vertical dashed line indicates the location of the LSO at $x=1/6$.

It is interesting to ask how reliable is the 5PN-accurate EOB-resummed 
analytical representation of the flux over the sequence of unstable orbits. 
We recall that Ref.~\cite{Damour:2008gu} introduced a specific factorization
and resummation of the PN waveform such that the related 
analytical flux was found to agree very well 
with the numerical one (see Fig.~1 (d) of~\cite{Damour:2008gu}). 
For this reason the top panel of Fig.~\ref{fig:hatf_ciro} additionally shows 
the energy flux constructed analytically from the resummed circularized multipolar 
waveform of~\cite{Damour:2008gu} that includes all the 5PN-accurate terms computed 
in~\cite{Fujita:2010xj}. The relative difference between fluxes is plotted in 
the bottom panel. The figure indicates a remarkable agreement between the analytical
and numerical fluxes also for circular orbits {\it below} the LSO, with a relative difference
that is almost always below $5\%$. Note that the difference becomes as large as
$10\%$ only for the last 6-7 orbits, which are very close to the 
light ring ($x=1/3$).
It is, however, remarkable that the analytical expression for the flux,
based on suitably resummed 5PN-accurate (only) results remains rather 
reliable in a region were the velocity of the orbiting particle is
about half the speed of light. It will be interesting in the future to
perform such a comparison with the 14PN-accurate expression of the 
waveform recently computed analytically by Fujita~\cite{Fujita:2011zk}.

In the spirit of the factorized form of the multipolar waveform entering the 
analytical flux, Eqs.~\eqref{eq:Flux}-\eqref{eq:multipoles}, the most important
information one wants to extract from the numerical data is  the behavior of the
residual amplitudes $\rho_\lm^{\rm Exact}(x)$ also along unstable orbits.
These quantities are the real unknowns of the problem, since all other factors, 
i.e.~the source $S^{(\epsilon)}(x)$ and the tail factor $T_{\ell m}(x)$, 
are known analytically. In this respect, the complete knowledge of the 
$\rho_\lm^{\rm Exact}$'s brings in the full strong-field information that is 
only partially available via their PN expansion 
The computation of $\rho_\lm^{\rm Exact}$ was performed for the first time in 
Ref.~\cite{Damour:2008gu}. It was restricted mainly to stable orbits, with
multipoles up to $\ell_{\rm max}=6$, and was based on the numerical data 
computed by Emanuele Berti~\cite{Pons:2001xs,Yunes:2008tw}. In addition, 
as mentioned above, a small sample of unstable orbits were also considered 
to explore the behavior of $\rho_{22}^{\rm Exact}$ toward the light ring. 

\begin{figure}[ht]
\center
\includegraphics[width=0.5\textwidth]{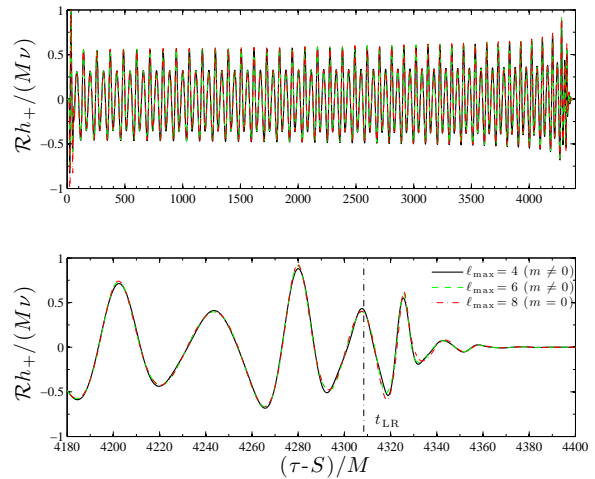}
\caption{\label{fig:htotal}(Color online). 
The $\R h_+/(M\nu)$ polarization (from Eq.~\eqref{eq:hplus_cross}) 
of the gravitational waveform for $\nu=10^{-3}$. 
The top panel shows the complete wave train ($\sim 40$ orbits up to merger). 
The bottom panel focuses around the merger time and illustrates the impact of 
subdominant multipoles. The vertical dashed line 
indicates the light-ring crossing time by the point-particle.} 
\end{figure}

The exact $\rho_\lm^{\rm Exact}$ are obtained from the partial fluxes $F_\lm^{\rm Exact}$ as
\be
\rho_\lm^{{\rm Exact},(\epsilon)}(x)=\left\{\dfrac{\sqrt{F_\lm^{\rm
      Exact}/F_\lm^{\rm Newton}}}{|T_\lm|
  \hat{S}^{(\epsilon)}}\right\}^{1/\ell} 
\ee
where the source $S^{(\epsilon)}$ is either the energy (for even-parity multipoles, $\epsilon=0$), 
or the Newton-normalized angular momentum (for odd-parity multipoles, $\epsilon=1$) 
along circular orbits, i.e.
\begin{align}
\hat{S}^{(0)}(x) &= \dfrac{1-2x}{\sqrt{1-3x}}\\
\hat{S}^{(1)}(x) &= \dfrac{1}{\sqrt{1-3x}} .
\end{align}
The square modulus of the tail factor $T_\lm$ reads~\cite{Damour:2007xr,Damour:2008gu}
\be
|T_\lm|^2=\dfrac{1}{(\ell!)^2}
\dfrac{4\pi\hat{\hat{k}}}{1-e^{-4\pi\hat{\hat{k}}}}\prod_{s=1}^\ell
\left[s^2 + \left(2\hat{\hat{k}}\right)^2\right]
\ee
where $\hat{\hat{k}}=m x^{3/2}$.

The result of the computation is presented in Fig.~\ref{fig:rholm_ciro} 
including multipoles up to $\ell_{\rm max}=7$. The figure clearly shows that, 
for some multipoles, the quasi-linear behavior of the $\rho_{\ell m}(x)$ 
above the LSO (explained in detail in~\cite{Damour:2008gu}) 
is replaced by a more complicated shape below the LSO, 
where high-order corrections seem relevant. 
The figure completes below the LSO the data of Fig.~3 
of~\cite{Damour:2008gu}, where only stable orbits were considered. 
Indeed, in the stable branch, the curves presented here 
perfectly overlap with those of~\cite{Damour:2008gu}.

We postpone to future work the analytical understanding of the behavior 
of the various $\rho_{lm}^{\rm Exact}$ when $x\to 1/3$. On the 
basis of the analytical information already contained in Fig.~5 of 
Ref.~\cite{Damour:2008gu}, it seems unlikely that the current 5PN-accurate 
analytical knowledge of the $\rho_\lm(x)$ functions can by itself 
explain the structure of the $\rho_{lm}^{\rm Exact}$ close to the 
light-ring. It will be interesting to see whether this structure
can be fully accounted for by the 14PN-accurate results 
of Ref.~\cite{Fujita:2011zk}.

\subsection{Gravitational radiation from inspiral, plunge, merger and ringdown}
\label{sbsec:inspl}

Now we discuss the properties of the gravitational wave
signal emitted by the five binaries with $\nu=10^{-2,-3,-4,-5,-6}$.
The initial relative separation is $r_0=7$ for $\nu=10^{-2,-3,-4}$,
$r_0=6.3$ for $\nu=10^{-5}$ and $r_0=6.1$ for $\nu=10^{-6}$. These 
latter values are chosen so that the evolution time is approximately
equally long for $\nu=10^{-4,-5,-6}$ ($\sim 400$ inspiral orbits, see Table~\ref{tab:recoil}).
The relative dynamics is started using post-circular initial data as described 
in~\cite{Buonanno:2000ef,Nagar:2006xv}, assuring a negligible initial
amount of eccentricity. The system is then driven by radiation reaction,
Eq.~\eqref{eq:rr}, into a (long) quasi-adiabatic inspiral, which 
is then smoothly followed by the nonadiabatic plunge phase, which
terminates with the merger of the two bodies and the final ringdown.
The relative dynamics and the multipolar structure of the waveforms 
are qualitatively the same as described in Paper I and II. 

\begin{figure*}[t]
\center
\includegraphics[width=0.45\textwidth]{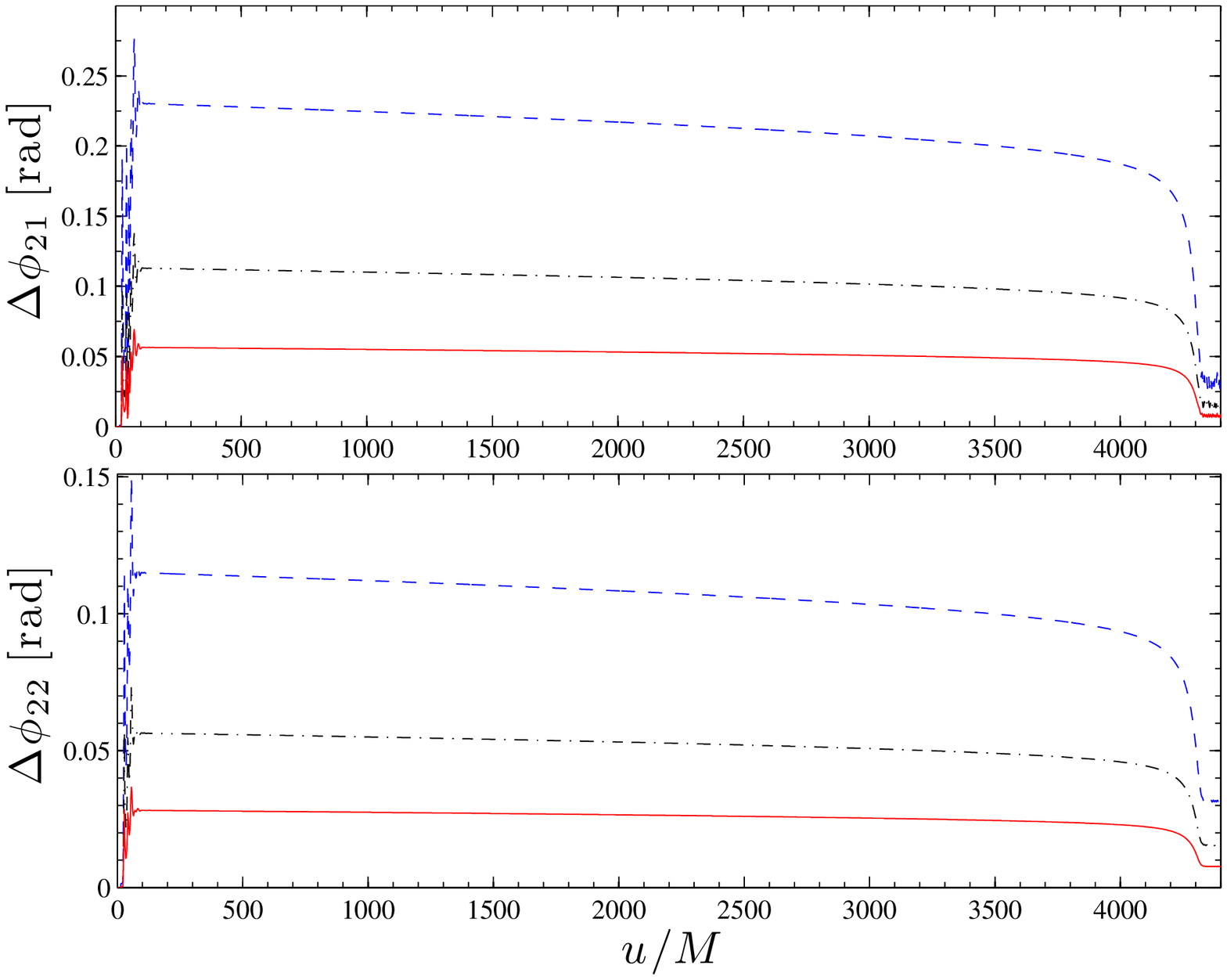}
\includegraphics[width=0.45\textwidth]{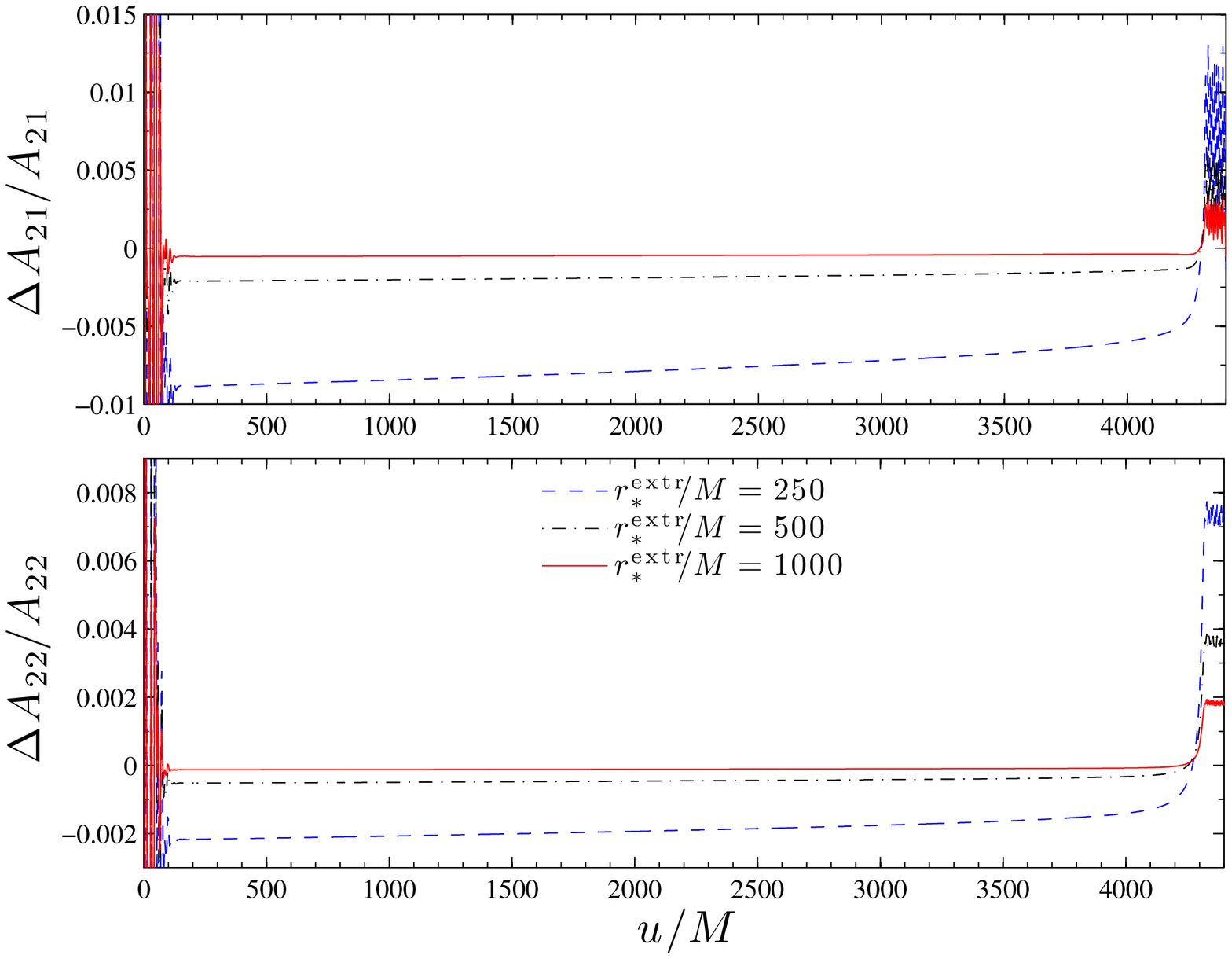}\\
\vspace{-0.1 in}
\includegraphics[width=0.45\textwidth]{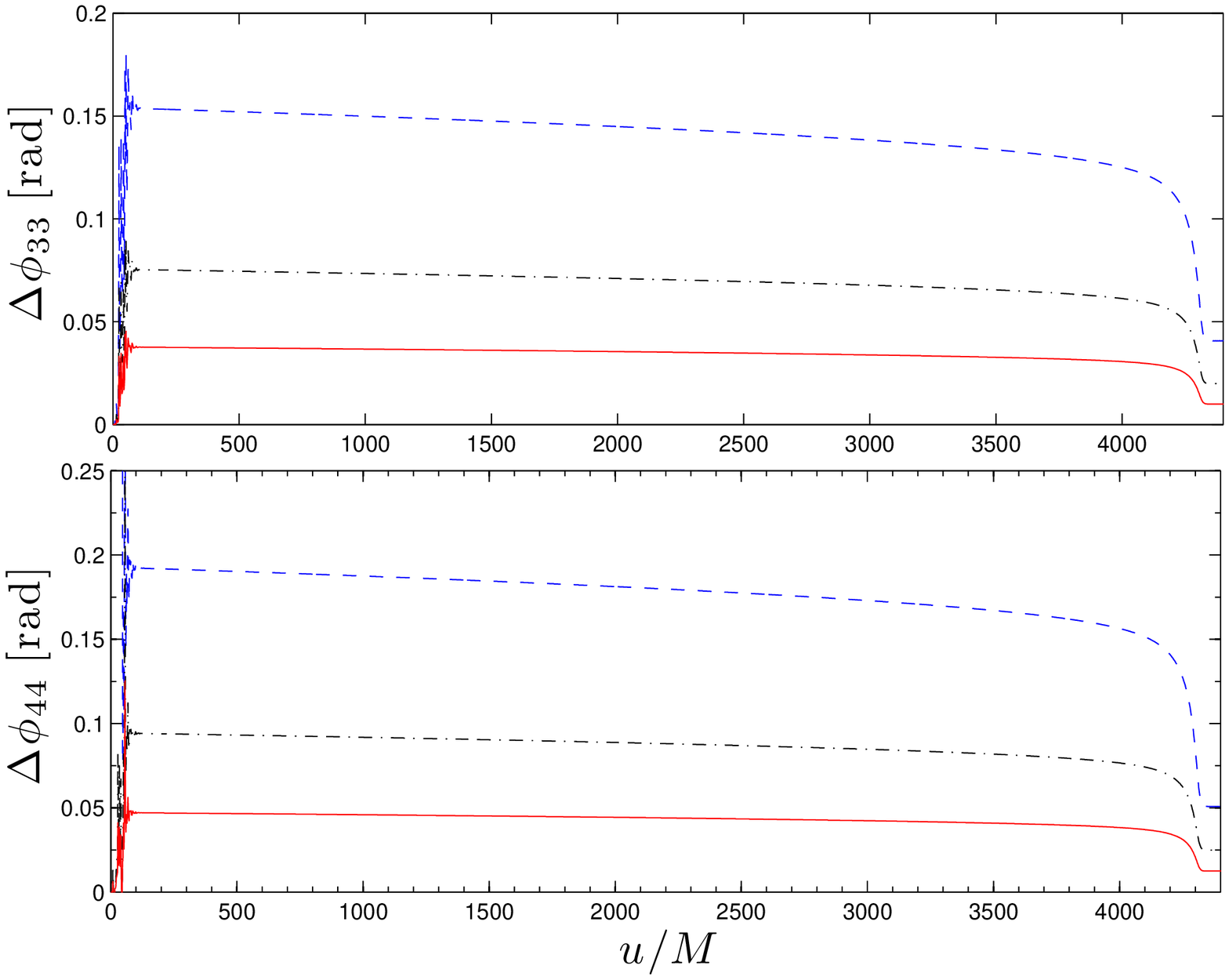}
\includegraphics[width=0.45\textwidth]{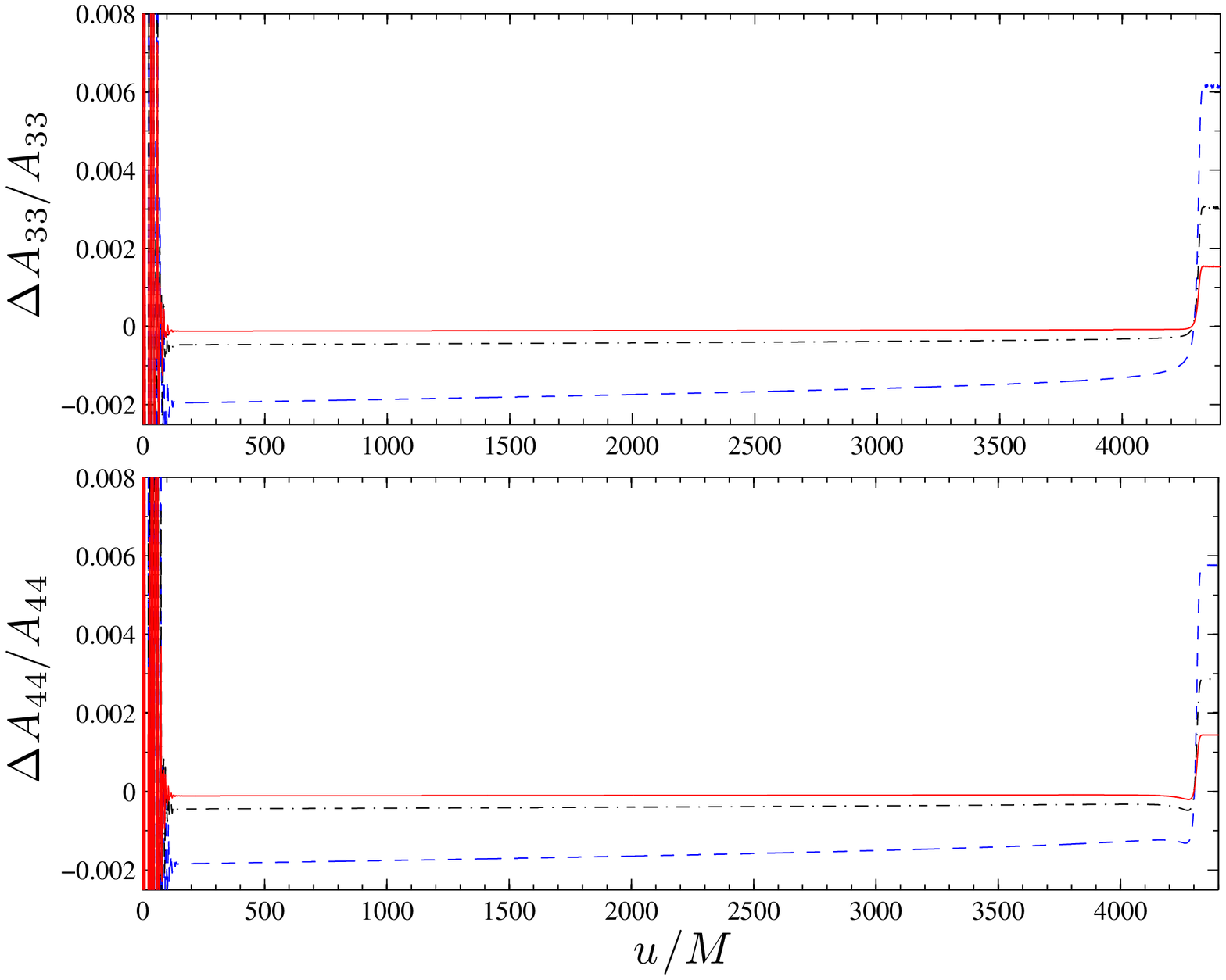}
\caption{\label{fig:pa_diff_lm}(Color online) Phase difference (left panels) 
and relative amplitude difference (right panels) between multipoles 
extracted at $\scri^+$ and at finite radii. Extraction radii are
$r_*^{\rm extr}/M=(250,\, 500,\, 1000)$.  
Data refer to the $\nu=10^{-3}$ binary.}
\end{figure*}

Let us discuss the mass ratio $\nu=10^{-3}$ as case study. 
We counted about $40$ orbits up to merger\footnote{With a slight 
abuse of definition, we consider the number of ``orbits'' 
as the value of the orbital phase at the end of the 
dynamical evolution divided by $2\pi$. In doing so we are also 
including in the computation the plunge phase, where the 
dynamics is nonadiabatic and cannot be approximated by 
a sequence of circular orbits.}, defined as the
time at which the particle crosses the light-ring ($r=3$).
Figure~\ref{fig:htotal} 
(displayed also in Paper II and Ref.~\cite{Sundararajan:2010sr})
shows the $\R h_+/(M\nu)$ polarization, Eq.~\eqref{eq:hplus_cross}, 
of the gravitational waveform for this binary along the fiducial direction 
$(\theta,\varphi)=(\pi/4,0)$ for various multipolar approximation.
The waveforms are displayed versus retarded time at $\scri^+$, 
$\tau - S$. The most accurate waveform includes the multipoles 
up to $\ell_{\max}=8$ (dash-dotted line). 
Summing up  to $\ell_{\rm max}=4$ captures most of the behavior up
to the light ring crossing ($t_{\rm LR}$, vertical dashed line), 
while the higher multipoles are more relevant during the 
late-plunge phase and ringdown. Note also the importance of 
the $m=0$ modes during the ringdown.

\subsubsection{Comparing waves extracted at $\scri^+$ and at finite radii}
\label{sbsc:finitextr}

Access to the radiation at $\scri^+$ enables us to evaluate 
finite distance effects in the waveform phase and amplitude.
We work again with mass ratio $\nu=10^{-3}$ only and compare 
waves extracted at $\scri^+$ with those extracted 
at three large, but finite, extraction radii 
$r_*^{\rm extr}/M=(250,\, 500,\,1000)$.
Figure~\ref{fig:pa_diff_lm} displays the phase differences 
$\Delta\phi_{\lm}\equiv\phi^{\scri^+}_{\lm}-\phi^{r^{\rm extr}_*}_{\lm}$  
(left panels) and the fractional amplitude 
difference $\Delta A_{\lm}/A_\lm\equiv(A^{\scri^+}_{\lm}-A^{r^{\rm extr}_*}_{\lm})/A_\lm^{\scri^+}$  
(right panels) for the most relevant multipoles.
On average, the phase differences accumulated between waves 
at $r_*^{\rm extr}/M=250 $ and at $\scri^+$ is $\Delta\phi_{\lm}\sim 0.125-0.25$~rad,
which decreases to $\Delta\phi_\lm\sim0.05$~rad when $r_*^{\rm extr}/M=1000$.
The corresponding fractional variation of the amplitude is
$\Delta A_\lm/A_\lm\sim0.2\%$ for $r_*^{\rm extr}/M=250$, which 
drops down by roughly a factor of $10$ for $r_*^{\rm extr}/M=1000$.
The phase differences shown in Fig.~\ref{fig:pa_diff_lm}
are \emph{significant}, in that they are much larger than the
numerical uncertainty ($\delta\phi\sim 10^{-6}$; see convergence results 
in Appendix~\ref{app:conv}).

\begin{figure}[ht]
\center
\includegraphics[width=0.46\textwidth]{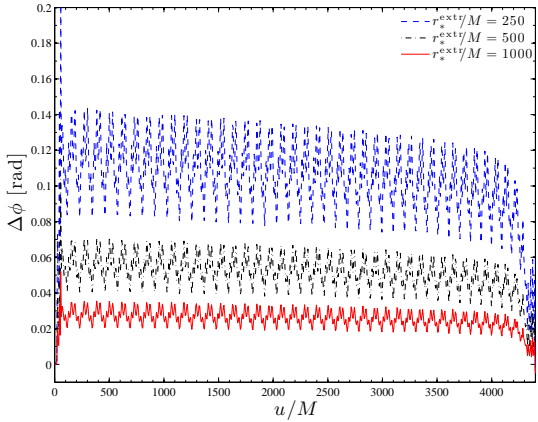}
\caption{\label{fig:phasediffh} (Color online) Phase difference between 
  the $\R h_+/(M\nu)$ total gravitational wave polarization at $\scri^+$ 
  and at finite radii. Data refer to the $\nu=10^{-3}$ binary.}
\end{figure}

An interesting feature that is common to both the phase difference 
and the fractional amplitude difference is that their variation is 
rather small during the inspiral, then decreases abruptly during the plunge 
(the LSO crossing is at $t_{\rm LSO}=u=4076.1$ for this binary) 
and the smallest values are reached during the ringdown. 
The multipolar behavior of Fig.~\ref{fig:pa_diff_lm} carries over 
to the total gravitational waveform. Figure~\ref{fig:phasediffh} shows
the phase difference between the total polarization $\R h_+/(M\nu)$
extracted at $\scri^+$ and at finite radii.
The phase difference amounts to (on average) $\Delta\phi\sim0.125$~rad 
for $r_*^{\rm extr}/M=250$ and $\Delta\phi\sim0.025$~rad for $r_*^{\rm extr}/M=1000$.
Note that the modulation in the phase difference is {\it not} numerical noise, 
but it is an actual physical feature due to the combination of the (different)
dephasings of the various multipoles.

We finally note that our $\ell=m=2$ EMR results are consistent 
with the corresponding equal-mass results displayed in Fig.~10 
of Ref.~\cite{Scheel:2008rj}, where they compare the extrapolated 
waveform to the one extracted at $r^{\rm extr}/M=225$.
After applying both a time and a phase shift to the finite-radius waveform,
they found that the accumulated phase difference to the extrapolated waveform
is of order $0.2$~rad, i.e.~about two times our (average)
dephasing for the $r_*^{\rm extr}/M=250$ waveform.

\subsubsection{Extrapolating finite-radius waveforms to $r\to\infty$}
\label{sbsc:extrap}

\begin{figure*}[t]
  \center
  \includegraphics[width=0.46\textwidth]{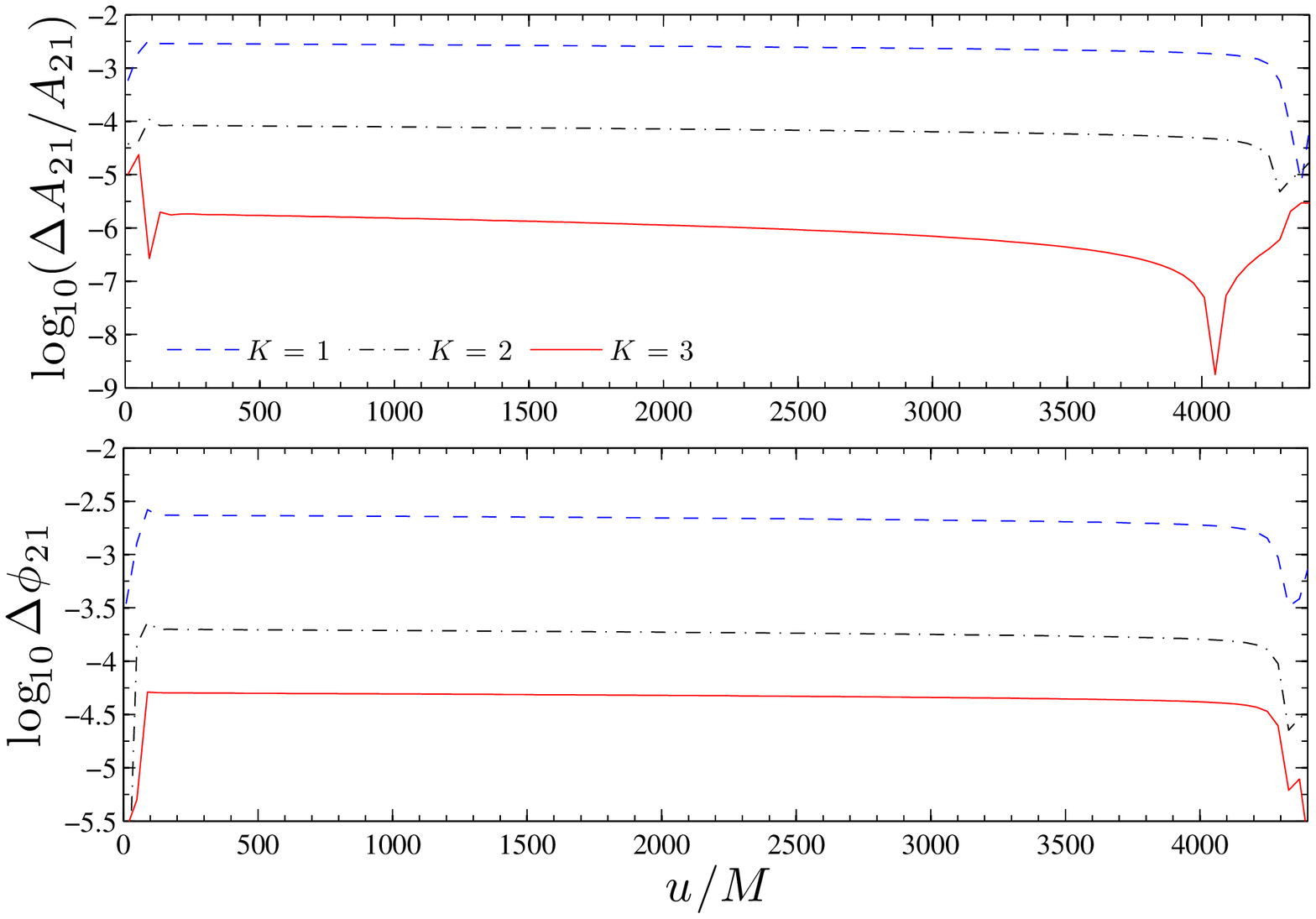}
  \includegraphics[width=0.46\textwidth]{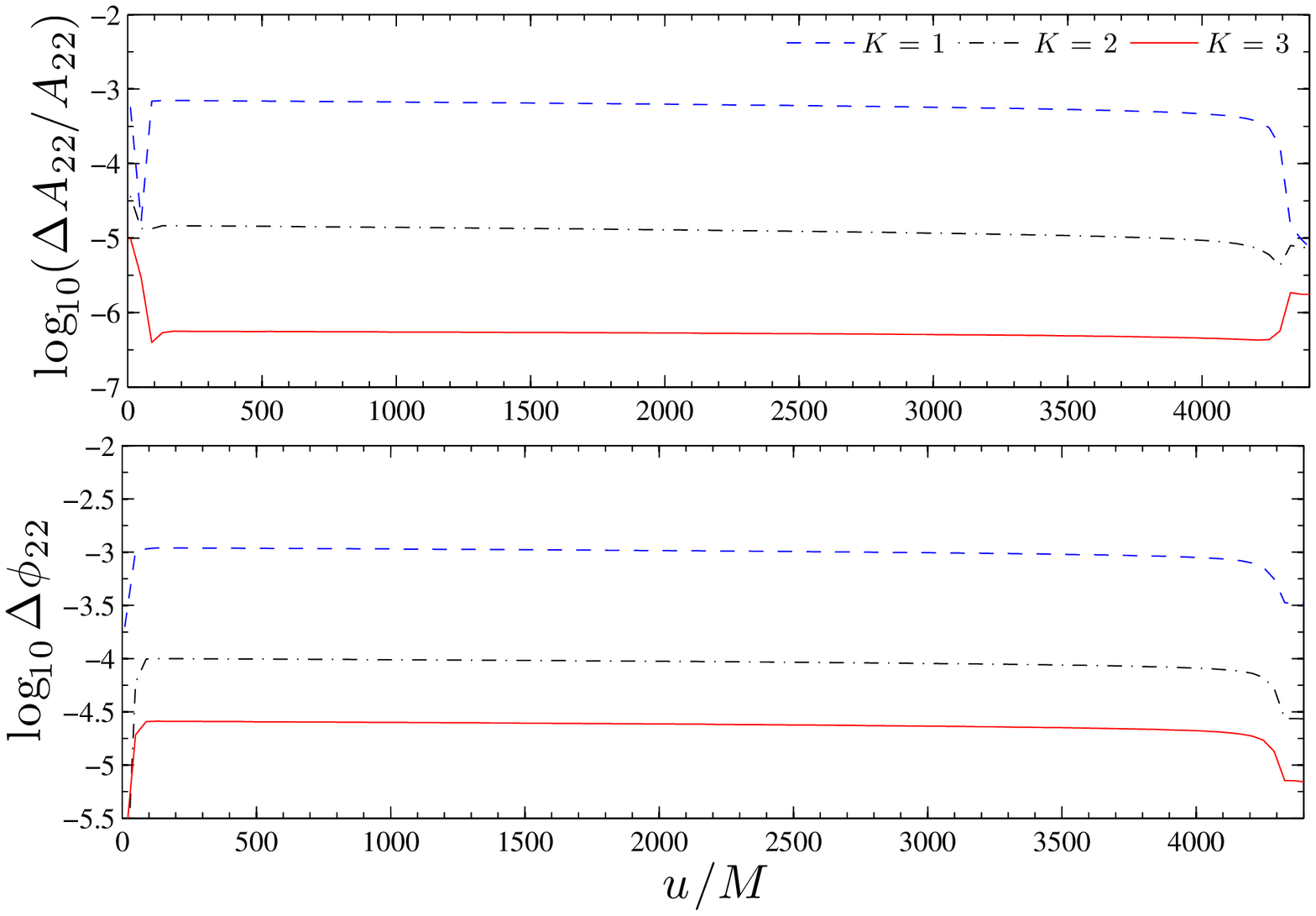}
  \caption{\label{fig:extrap} (Color online). Fractional amplitude difference 
    (top panels) and dephasing (bottom panel) between $\scri^+$ and extrapolated
    waveforms. Note that we plot the $\log_{10}$. Multipoles are 
    $\ell=m=2$ (left panels) and  $\ell=2$, $m=1$ (right panels). 
    Extraction radii are $r_*/M\simeq(250,500,750,1000)$. Different 
    lines refer to different polynomial order in the extrapolation i.e. $K$ in
    Eq.~\eqref{eq:extrap}. The plot refers to the $\nu=10^{-3}$
    binary.}
\end{figure*}

Now that we have shown that finite-radius effects are significant,
we use the data at $\scri^+$ to test, in a well controllable
setup, the standard extrapolation to $r\to\infty$ routinely
applied to NR finite-radius waveforms.

Indicating with $r$ the radius
at which radiation is measured in NR simulations, 
the waveforms are extrapolated to \mbox{$r\to\infty$}
by assuming an expansion in powers of $1/r$ 
(see e.g. Refs.~\cite{Boyle:2007ft,Scheel:2008rj,Pollney:2009ut,Boyle:2009vi}), 
\be
\label{eq:extrap}
f(u,r) = \sum_{k=0}^{K}\frac{f_k(u)}{r^k} \ ,
\ee
where $f$ can be either the amplitude or the phase of the
gravitational waveform\footnote{ %
  In NR studies the extrapolation is usually applied to the curvature 
  waveform $r\,\psi_4$.
}.
The extrapolation procedure of NR data is affected by the fictitious 
identification of a background (Schwarzschild or Kerr) in the numerically 
generated spacetime and by subtleties in the definition of the retarted 
time for each observer (see e.g. Sec.~IIB of Ref.~\cite{Boyle:2009vi} 
and Sec.~IIIC of Ref.~\cite{Scheel:2008rj}). 
Thanks to the aforementioned CCE procedure to compute the GW signal 
at $\scri^+$, Ref.~\cite{Reisswig:2009us} was able to provide an independent 
check of the extrapolation procedure. Reference~\cite{Reisswig:2009us}
focused on the $\ell=m=2$ $\psi_4$ waveform from an equal-mass 
black-hole binary and considered data extracted at 
$r/M=(280,300,400,500,600,1000)$ as  input 
for the extrapolation procedure. Over the $1000M$ of evolution from early inspiral to
ringdown, Ref.~\cite{Reisswig:2009us} found a dephasing of 0.019 rad and a maximum 
fractional amplitude difference of $1.08\%$ between the extrapolated and 
the $\scri^+$ waveforms. 

Our setup permits the validation of the expansion in Eq.~\eqref{eq:extrap} 
and a quantification of the extrapolation errors in the absence of 
ambiguities related to the definition of the extraction spheres
and retarded times  on a dynamical spacetime. 
The radius, $r$, is the areal 
radius of the Schwarzschild background and the retarded time is by construction
$u=\tau-\rho$.
To produce a meaningful comparison with the estimates of~\cite{Reisswig:2009us},
we use waveforms extracted at $r_*^{\rm extr}/M=(250,500,750,1000)$
as input for the extrapolation procedure, and
we work again with the $\nu=10^{-3}$ binary.  

\begin{figure}[t]
  \center
  \includegraphics[width=0.44\textwidth]{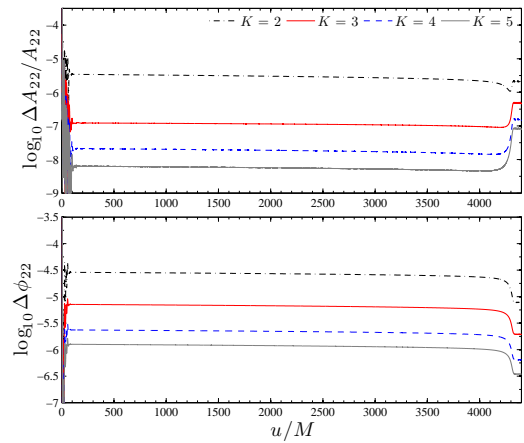}
  \caption{\label{fig:extrap_many}(Color online). Residual of amplitude (top) 
    and phase (bottom) between the $\scri^+$ and the extrapolated $\ell=m=2$ waveform. 
    Note that we plot the $\log_{10}$.
    Extraction radii are $r_*^{\rm extr}/M= (250,\,500,\,750,\,1000,\,2000,\,4000)$. 
    Different lines refer to different polynomial order in the extrapolation 
    i.e. different $K$ in Eq.~\eqref{eq:extrap}. Data refer to $\nu=10^{-3}$ binary.}
\end{figure}

The phase and amplitude differences are plotted in Fig.~\ref{fig:extrap},
where we show only $\ell=2$ multipoles for definitess (the picture does
not change for other multipoles): 
$m=1$ (left panel) and $m=2$ (right panel).  Different lines in the 
plot correspond to different choices of the maximum power $K$ 
in the polynomial expansion~\eqref{eq:extrap}.
The phase difference between the wave at $\scri^+$ 
and the extrapolated one decreases uniformly in time: It is between 
$10^{-2}$ and $10^{-3}$ rad when a linear polynomial ($K=1$) in $1/r$ 
is assumed in Eq.~\eqref{eq:extrap} and it drops to between $10^{-4}$ 
and $10^{-5}$ when a cubic polynomial is used ($K=3$). 
In this analysis we considered only up to $K=3$ because 
this value seems to give the best compromise between noise and accuracy 
when extrapolating NR waveforms~\cite{Scheel:2008rj,Pollney:2009ut}. 
We remark, however, that in our setup we are not limited in the choice of $K$. 
This is evident in Fig.~\ref{fig:extrap_many} where we use higher values of $K$ and
more extraction radii $r_*^{\rm extr}/M=(250,\, 500,\, 750,\,1000,\, 2000,\,4000)$,
for the $\ell=m=2$ waveform. Both the phase and amplitude differences 
decrease monotonically with increasing $K$, 
showing that more powers in 
the expansion~\eqref{eq:extrap} lead to more accurate extrapolation.
The simple extrapolation formula~\eqref{eq:extrap} proves
robust and leads to reliable waveforms. 

\subsection{Angular momentum loss}
\label{sbsec:angmom}

\begin{figure}[ht]
\center
\includegraphics[width=0.52\textwidth]{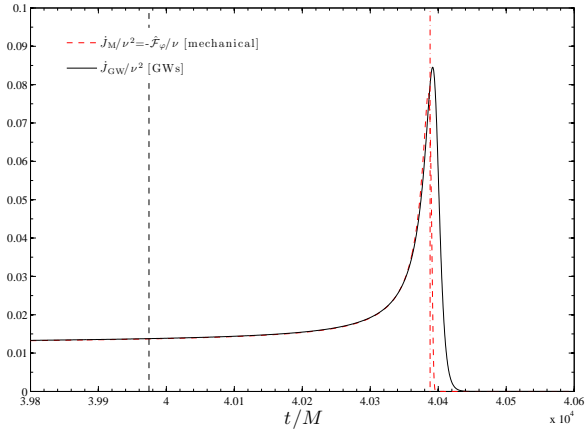}
\caption{\label{fig:flux_comp}(Color online). Late-time comparison between two angular momentum 
losses for the binary with $\nu=10^{-4}$. The GW flux ($\dot{J}_{\rm GW}/\nu^2$, solid line) 
computed from the RWZ waveform and extracted  at $\scri^+$ (including
up to $\ell_{\rm max}=8$ radiation  
multipoles) is contrasted with the EOB-resummed, analytical mechanical angular momentum  
loss $-\hat{\F}/\nu$ (dashed line). 
The two vertical lines correspond (from left to right) to the particle crossing respectively, 
the adiabatic LSO location ($r=6$, $t_{\rm LSO}=39974.40$), and the light-ring location 
($r=3$, $t_{\rm LR}=40388$).} 
\end{figure}

\begin{figure}[ht]
\center
\includegraphics[width=0.5\textwidth]{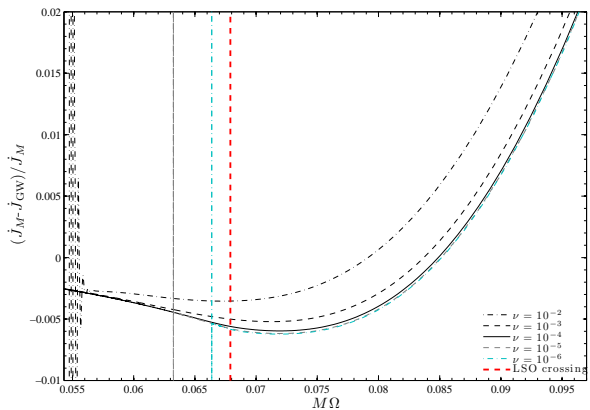}
\caption{\label{fig:DJ_Omg}(Color online). Relative difference between the 
mechanical angular momentum loss and the GW energy flux for the five mass 
ratios considered. The figure highlights how a very small fractional 
difference is maintained also after the LSO crossing.} 
\end{figure}

The main uncertainty in our approach lies, as discussed above, on
the accuracy of the analytically resummed radiation reaction, 
Eq.~\eqref{eq:rr}. Several studies~\cite{Damour:2007xr,Bernuzzi:2010ty} 
have shown the {\it consistency} between the gravitational wave angular momentum 
flux computed from the RWZ waveform (measured at a large, finite radius) 
and the mechanical angular momentum loss $-\hat{\F}_\varphi$ obtained 
by suitably resumming (a la Pad\'e) the Taylor-expanded 
PN flux~\cite{Damour:2007xr}, or via the multiplicative decomposition
of the waveform of~\cite{Damour:2007xr,Damour:2008gu,Fujita:2010xj}, 
as performed in~\cite{Bernuzzi:2010ty}. 
In particular, Ref.~\cite{Bernuzzi:2010ty} pointed out a fractional difference 
between mechanical and GW angular momentum fluxes at the $10^{-3}$ level up
to (and even below) the adiabatic LSO crossing. The 
common drawback of these studies is that the target ``exact'' flux is computed
at a finite extraction radius (typically $r_*/M=1000$), whereas the analytical 
$\cal{F}_\varphi$ is computed (by construction) at $\scri^+$.
Because we can compute the RWZ flux at $\scri^+$, 
the comparison between the instantaneous GW angular momentum
flux $\dot{J}_{\rm GW}/\nu^2$ and the mechanical angular momentum loss 
$\dot{J}_M/\nu^2=-{\F}_\varphi/\nu$ is more meaningful, and can be 
calculated without the ambiguity caused by a relative time-shift that one 
should include when $\dot{J}/\nu^2$ is computed at a finite radius
(it was not included in~\cite{Bernuzzi:2010ty} for simplicity).

We focus first on the $\nu=10^{-4}$ simulation.
In Fig.~\ref{fig:flux_comp}  we compare the mechanical angular
momentum loss (changed sign, $-\hat{\F}_\varphi/\nu$, dashed line)  to the 
instantaneous angular momentum flux ($\dot{J}_{\rm GW}/\nu^2$, solid line)
extracted at $\scri^+$ and plotted versus the corresponding retarded 
time $\tau-S$. Since $\hat{F}_\varphi$ is parametrized by the mechanical
time $t$, we use this as $x$-axis label.
The two vertical lines on the figure indicate (from left to right) 
the particle crossing of the adiabatic LSO location  
($r/M=6$, $t_{\rm LSO}=39974.40$, dashed black line), 
which can be considered approximately as the end of the inspiral, 
and the light-ring crossing ($r/M=3$, $t_{\rm LR}=40388$, dashed red line).
Consistently with the findings of Paper I 
(compare Fig.~8 in Paper I, which used the flux at $r_*^{\rm extr}/M=1000$), 
the figures confirm visually the good agreement between the two fluxes 
also below the LSO crossing, and actually almost during the entire plunge phase. 
The relatively large difference between the fluxes around the light-ring crossing 
is due to the lack of 
next-to-quasi-circular (NQC) corrections in the waveform amplitude as well as of
ringdown quasi-normal-modes, in the analytically constructed $\dot{J}_M/\nu^2$. 
Note that Paper II has explicitly shown how these 
corrections can be effectively added to the ``bare'' inspiral resummed 
multipolar waveform that we use to compute radiation reaction to obtain
a much closer agreement between the waveform moduli in the 
strong-field-fast-velocity regime. 
We work with NQC-free radiation reaction because the late part of 
the dynamics (and waveform) is practically unaffected by details 
of the radiation reaction, as discussed in~\cite{Damour:2007xr}.

The qualitative agreement seen in Fig.~\ref{fig:flux_comp} is depicted more 
accurately in Fig.~\ref{fig:DJ_Omg}. The figure displays (for the five mass
ratios considered) the relative difference $(\dot{J}_M-\dot{J}_{\rm GW})/\dot{J}_M$
versus the orbital frequency $M\Omega$. For reference, the LSO crossing
frequency, $M\Omega_{\rm LSO}\approx 0.068$, is marked by a vertical dashed line 
(red online) in the figure\footnote{Note that the other two apparent vertical 
lines are actually the junk radiation corresponding to the beginning of the 
$\nu=10^{-5}$ and $\nu=10^{-6}$ simulations.}. 
For $\nu=10^{-3}$, the relative difference is initially at $2.5\times 10^{-3}$
and then it slowly increases to reach only $5\times 10^{-3}$ 
at the LSO crossing. These (rather small) differences are due to the limited 
PN knowledge (5PN) at which the residual multipolar amplitudes $\rho_{\ell m}$ 
are implemented in the radiation reaction.
When considering $\nu=10^{-4}$, still starting at $r_0=7$, the picture remains
practically unchanged (solid line in the figure), although the difference
is slightly larger at the LSO crossing and during the plunge. The cases of $\nu=10^{-5}$
and $\nu=10^{-6}$ (that start respectively at $r_0=6.3$ and $r_0=6.1$) 
are practically superposed and one sees again a slight increase
of the difference around the LSO. This agreement is a strong indication that the 
analytically resummed radiation-reaction force is suitable to drive the dynamics 
of a  (circularized) EMRI, notably with $\nu=10^{-6}$, an interesting source 
for LISA\footnote{A similar conclusion was also reached in Refs.~\cite{Yunes:2009ef,Yunes:2010zj}, 
that actually pointed out that one should properly calibrate the $\F_\varphi$ function
to have an accurate representation of the EMRI dynamics. Note however that here,
contrarily to Refs.~\cite{Yunes:2009ef,Yunes:2010zj}, we include in the discussion also
the late inspiral and plunge regime.}. In the future, it should be explored how this 
agreement improves when the 14PN-accurate corrections to the 
$\rho_{\lm}$ from ~\cite{Fujita:2011zk} are included in the flux. 
 
As a last remark, Fig.~\ref{fig:DJ_Omg} also highlights that the differences between 
the various curves become smaller and smaller when $\nu\to 0$. 
In particular, the curves for mass ratios $\nu=10^{-4,-5,-6}$ are almost superposed, 
which points out that radiation reaction has little effect during the 
plunge phase for these binaries. This fact suggests that, when $\nu \lesssim 10^{-4}$, 
the motion is ``quasi-geodesic'' around and below LSO crossing~\cite{Bernuzzi:2010ty}, 
i.e., it is a good approximation to the geodesic plunge from the LSO~\cite{Damour:2007xr,Hadar:2009ip,Hadar:2011vj}.

\subsection{Gravitational recoil}
\label{sbsc:kick}
We update the calculation of the ($\nu\to 0$) recoil velocity performed in Paper 
I (see also Ref.~\cite{Sundararajan:2010sr}) using the new data 
extracted at $\scri^+$ and considering more mass ratios. 
The linear momentum flux emitted in GWs is computed via Eq.~\eqref{eq:dPdt} 
(with $\ell_{\rm max}=7$) and is integrated in time to obtain the  accumulated 
complex recoil velocity as
\be
v\equiv v_x+\ii v_y = v_0 - \dfrac{1}{M}\int_{t_0}^t\left( {\cal F}^{\bf P}_x +\ii {\cal F}^{\bf P}_y \right)dt.
\ee
Here, $t_0$ is the initial time of the simulation and $v_0$ is the initial
velocity that the system has acquired from $t=-\infty$ to $t=t_0$. 
We give a good approximation to $v_0$ as in Sec.~IV of Paper~I 
(and of Ref.~\cite{Pollney:2007ss}), i.e., by determining the center 
of the hodograph  (see Fig.~5 of Paper~I) of the complex recoil 
velocity during part of the inspiral.

Table~\ref{tab:recoil} lists both the (modulus of) the maximum and the 
final kick velocity for the five mass ratios considered, together with the total number of
orbits, $N^{\rm orbits}$, and the number of orbits used to determine $v_0$,
$N^{\rm orbits}_0$. The uncertainty in the numbers is on the last digit 
(of order $\times 10^{-5}$) and is estimated from the variation of
$|v^{\rm end}|$ and $|v^{\rm max}|$ when $N^{\rm orbits}_0$ is 
modified\footnote{The perturbative treatment is not meant to give an accurate 
estimate of the final recoil for $\nu=10^{-2}$, because 
high-order, $\nu$-dependent corrections in the dynamics (and waveforms) 
are important in this case. In fact, the NR simulation of 
Ref.~\cite{Lousto:2010ut} gives $|v^{\rm end}|/\nu^2 = 0.037\pm 0.002$, 
which is $17\%$ smaller than the perturbative estimate. 
Nonetheless, the result of Ref.~\cite{Lousto:2010ut} is consistent 
with the fit analysis of Fig.~7 of Paper~I.}.
Note that the values are slightly larger than those of Table~III of 
Paper~I, which were measured at $r_*^{\rm extr}/M=1000$.
Because, as observed before, the dynamics is practically independent 
on $\nu$ for $\nu\leq 10^{-4}$, i.e.~the motion is quasi-geodesic, 
we can average the results for $\nu=10^{-4,-5,-6}$ in Table~\ref{tab:recoil} 
so to obtain 
an estimate of the final and maximum recoil in the $\nu=0$ case,
with an uncertainty given by the corresponding standard deviation.
This calculation gives $v_{\rm kick}^{\rm end}/(c \nu^2)=0.04474\pm 0.00007$ and
$v_{\rm kick}^{\rm max}/(c\nu^2)=0.05248\pm 0.00008$.

\begin{table}[t]
  \caption{\label{tab:recoil} Computation of the kick velocity.
    From left to right, the columns report: the mass ratio $\nu$; the
    initial separation $r_0$; the total number of orbits, the number
    of orbits used to determine an approximate value of the correct
    initial kick velocity $v_0$ (see Sec.~IV of Paper~I); 
    the final kick velocity $|v^{\rm end}|$ and the maximum kick velocity $|v^{\rm max}|$. 
  }
\begin{center}
\begin{ruledtabular}
\begin{tabular}{cccccc}
$\nu$      & $r_0$ & $N^{\rm orbits}$ & $N_0^{\rm orbits}$ & $|v^{\rm end}|/(c\nu^2)$  & 
              $|v^{\rm max}|/(c\nu^2)$ \\
\hline
$10^{-2}$   & $7.0$  & 6   & 2 & 0.0435(6)  &  0.0508(1)  \\
$10^{-3}$   & $7.0$  & 40  & 14 & 0.0445(6)  &  0.0522(8)  \\
$10^{-4}$   & $7.0$  & 375 & 123 & 0.0447(6)  & 0.0525(3)  \\
$10^{-5}$   & $6.3$  & 349 & 115 & 0.0446(6)  & 0.0523(9)  \\
$10^{-6}$   & $6.1$  & 396 & 133 & 0.0448(0)  & 0.0525(2)   
\end{tabular}
\end{ruledtabular}
\end{center}
\end{table}

\section{Conclusions}

In this paper we discussed, for the first time, the hyperboloidal layer
method, introduced in Ref.~\cite{Zenginoglu:2010cq}, for the
computation of the gravitational radiation emitted by 
large-mass-ratio compact binaries at null infinity.
We used a hyperboloidal layer in a perturbative,
time-domain method specifically designed for computing EMR 
(or IMR) waveforms without the adiabatic assumption.
The method employs the RWZ formalism for wave generation
and an analytic, EOB-resummed, leading order, radiation reaction
for the dynamics of the 
particle~\cite{Nagar:2006xv,Damour:2007xr,Damour:2008gu,Fujita:2010xj,Bernuzzi:2010ty}.
Higher $\nu$-dependent conservative and nonconservative corrections
to the relative dynamics, as present in the complete EOB formalism,  
are neglected by construction. Merged with the hyperboloidal method, 
the method efficiently provides accurate waveforms at null infinity. 
These waveforms have already been used to calibrate effective
next-to-quasi-circular corrections to the multipolar EOB-waveform 
(amplitude and phase) in the test-mass limit~\cite{Bernuzzi:2010xj}. 
 
In this paper, beside providing an extensive discussion of the
hyperboloidal technique, we presented results concerning the study of
the gravitational 
radiation from circular stable and unstable orbits, and from the
coalescence of circularized black-hole binaries 
with mass ratios $\nu=10^{-2,-3,-4,-5,-6}$.
We improved quantitatively previous work~\cite{Nagar:2006xv,Damour:2007xr,
Bernuzzi:2010ty,Sundararajan:2010sr}, where waves were extracted 
at finite radii. The difference of the null infinity
waveforms to finite-radius and extrapolated waveforms are quantified in detail.
The waveforms produced in this work will be made publicly available
so to be used in data-analysis pipelines for LISA-type science  
or for the Einstein Telescope.
Below we discuss our results individually, together with an outlook.

\paragraph*{Circular orbits.} We computed the gravitational energy
flux emitted by a particle in geodesic circular motion.
We considered a sample of (strong field) circular orbits and found
that the flux agrees with the semi-analytic data of Fujita 
et al.~\cite{Fujita:2004rb} within at most a $0.8\%$ 
in each multipole. The total flux, summed up to $\ell_{\rm max}=8$,
agrees always within $0.02\%$ for all orbits up to the LSO, $r_0=6$. 
We considered also unstable circular orbits, $3.1\leq r_0<6$, which
are useful to test the performance of the waveform resummation 
procedure below the LSO~\cite{Damour:2008gu}. The Newton-normalized
energy flux computed within our approach (considered ``exact'' for this comparison) 
is compared with the homologous, EOB-resummed analytic expression. 
We found a relative difference always below $5\%$ until $r_0=4.2$ with a
maximum of $10\%$ at $r_0=3.3$.
We also computed from the numerical data the ``exact'' residual 
waveform amplitudes $\rho_\lm^{\rm Exact}$ introduced in~\cite{Damour:2008gu}, 
although we did not provide a thorough comparison with their 
5PN-accurate analytic counterparts~\cite{Fujita:2010xj}.

\paragraph*{High accuracy inspiral waveforms at null infinity.} 
We computed high accuracy waveforms covering the complete 
transition from a (long, $\sim 400$ orbits for 3 mass ratios) 
quasi-circular inspiral to plunge, merger and ringdown phases.
The phase and amplitude error bars on the dominant multipoles, 
as estimated from convergence tests, are $\delta\phi\sim 10^{-6}$ 
and $\delta A/A\sim 10^{-6}$. 
The multipolar structure of the gravitational wave is qualitatively the same 
as reported in Paper I and II.

\paragraph*{Self-consistency of the method.} The perturbative method
proposed here has systematic uncertainties in the assumption made for
the radiation reaction. To check the self-consistency of our method we 
compared the mechanical angular momentum loss and the angular momentum
flux computed from the waves. For $\nu\leq 10^{-3}$, the relative disagreement 
between the two is $\sim 2.5\times10^{-3}$ at the beginning of the simulations 
and reaches only $\sim 5\times10^{-3}$ at the LSO crossing, which is then maintained  
up to orbital frequency $M\Omega\sim0.085$. This agreement supports 
the reliability of the analytical resummed radiation reaction model.

\paragraph*{Comparison with finite radii extraction.}
We found significant differences in waveforms at finite-radii and at $\scri^+$. 
For example, the $\ell=m=2$ multipole extracted at $r^{\rm extr}_*/M=250$ 
differs from the $\scri^+$ waveform (on average) 
by $\Delta\phi_{22}\sim 0.1$~rad and
by $\Delta A_{22}/A_{22}\sim 0.2\%$ during the late-inspiral 
and plunge; the differences reduce to $\Delta\phi_{22}\sim 0.025$~rad
and $\Delta A_{22}/A_{22}\sim 0.01\%$ at $r^{\rm extr}_*/M=1000$. 
Such differences, though small, are relevant in the comparison 
with the EOB-resummed analytic waveform~\cite{Bernuzzi:2010xj}.

\paragraph*{Extrapolation to infinite extraction radius.}
We extrapolated the finite-radius waveforms to infinite radius 
using a simple $1/r$-polynomial expression, Eq.~\eqref{eq:extrap}, 
as routinely applied to NR waveforms. 
Considering the $\ell=m=2$ waveform, we found that the
dephasing between the extrapolated and the $\scri^+$ waveforms
reaches $10^{-3}$~rad using a linear polynomial 
in $1/r$ and extraction radii below $1000M$. In our setup, the dephasing can be
made small to the level of our uncertainties, 
by simply improving the extrapolation procedure: 
for example, it drops to $10^{-6}$ using larger radii (up to $4000M$) and higher powers
($K=5$) in the extrapolation. Note that we work within perturbation theory where
a fixed, explicitly spherically symmetric background is given. In NR, the
extrapolation procedure is not as well-defined due to various factors 
such as the (arbitrary) definition of a retarded time, the negligence of gauge
dynamics, the identification of a fiducial background, and the use of coordinate
spheres. Consequently, the dephasings can be several orders of magnitude larger
for NR waveforms of coalescing black-hole binaries~\cite{Scheel:2008rj,
Reisswig:2009us}. This observation emphasizes the importance of using
unambiguous extraction procedures to compute NR waveforms at $\scri^+$, 
such as the CCE implemented by Ref.~\cite{Reisswig:2009us}.

\paragraph*{Gravitational recoil.} 
We updated the final recoil computed in Paper~I 
and in Ref.~\cite{Sundararajan:2010sr}. We computed the maximum and
final kick for several mass ratios (see Table~\eqref{tab:recoil})
and then extrapolated these values to the $\nu\to 0$ limit. Our final 
estimates for the maximum and final recoil velocities are 
$v^{\rm max}_{\rm kick}/(c\nu^2)=0.05248\pm 0.00008$ 
and $v^{\rm end}_{\rm kick}/(c\nu^2)=0.04474\pm 0.00007$. 
These values can be used together with NR data to provide fitting
formulas valid for all values of $\nu$ (see Paper I).

\paragraph*{Outlook.}
An important development that is called for the future is the computation
of the gravitational radiation emitted in the coalescence of {\it noncircularized}
binary systems. In the IMR regime, these systems might be interesting
sources for the Einstein Telescope or for planned space interferometers.
Within the EOB approach, there are prescriptions~\cite{Damour:2004bz,Damour:2005ug}
to resum  the radiation reaction in the noncircularized case and to account for 
the radiation-reaction driven evolution of the eccentricity. 
Such prescriptions (in their $\nu=0$ limit) can be easily implemented 
in our framework and we plan to do it in a future study. 

It will be interesting to motivate {\it analytically} the
behavior of the residual (numerical) amplitudes $\rho_\lm^{\rm Exact}$ along 
the sequence of unstable circular orbits. To do so, it will be necessary 
to compare our results with the 14PN-accurate analytical expressions 
recently obtained by Fujita~\cite{Fujita:2011zk}. The 
assessment of the accuracy of the analytical $\rho_\lm$'s for unstable 
orbits might then be useful to study the dynamical modification to geodesic 
zoom-whirl orbits due to the action of (EOB-resummed) radiation reaction.

Such studies would also benefit from technical improvements of the finite-difference,
time-domain RWZ infrastructure. Possible improvements for the future are the 
implementation of horizon-penetrating coordinates to increase efficiency and remove
reflections from the inner boundary, a better implementation of the initial 
data that satisfies the linearized Einstein constraint equations, 
and the use of higher order finite-difference methods to reduce numerical truncation error.
One can also extend our numerical framework to spinning black holes by 
solving, instead of the RWZ equation, the Teukolsky equation for computing the
gravitational waves emitted by a small black hole inspiraling into a rotating black
hole. 

The waveforms produced in this work can be used as a benchmark 
for the consistency of NR results in the large mass ratio
limit~\cite{Lousto:2010ut,Sperhake:2011ik}.
  
Finally, we hope that this work will motivate future efforts towards a full 
understanding of the hyperboloidal initial value problem in nonlinear general
relativity, with particular attention to its application in numerical relativity.

\acknowledgements

We thank Thibault Damour for useful inputs. We are also grateful to 
Ryuichi Fujita for making available to us his data for circular orbits.
SB is supported by DFG Grant 
SFB/Transregio~7 ``Gravitational Wave Astronomy.''  SB thanks IHES for 
hospitality and support during the development of this work.
AZ acknowledges support by the NSF Grant No.~PHY-1068881, 
and by a Sherman Fairchild Foundation grant to Caltech.
Computations were performed on the {\tt MERLIN} cluster at IHES. 
The authors thank Francois Bachelier for computer assistance.

\appendix

\section{Convergence tests and errorbars}
\label{app:conv}
\begin{table}[ht]
  \caption{\label{tab:gwflux} Gravitational energy and angular
    momentum fluxes  at $\scri^+$ for a particle on a 
    circular orbit of radius $r_0=7.9456$. 
    Compare with Refs.~\cite{Martel:2003jj,Sopuerta:2005gz}.}
\begin{center}
\begin{ruledtabular}
\begin{tabular}{cccc}
  $\ell$ & $m$ & $\dot{E}/\nu^2$ & $\dot{J}/\nu^2$ \\
\hline
2& 1& $8.1632\times 10^{-7}$& $1.8283\times 10^{-5}$\\ 
2& 2& $1.7065\times 10^{-4}$& $3.8220\times 10^{-3}$\\ 
\hline
3& 1& $2.1740\times 10^{-9}$& $4.8691\times 10^{-8}$\\ 
3& 2& $2.5203\times 10^{-7}$& $5.6448\times 10^{-6}$\\ 
3& 3& $2.5481\times 10^{-5}$& $5.7070\times 10^{-4}$\\ 
\hline
4& 1& $8.4001\times 10^{-13}$& $1.8814\times 10^{-11}$\\ 
4& 2& $2.5112\times 10^{-9}$& $5.6243\times 10^{-8}$\\ 
4& 3& $5.7777\times 10^{-8}$& $1.2940\times 10^{-6}$\\ 
4& 4& $4.7289\times 10^{-6}$& $1.0591\times 10^{-4}$\\ 
\hline
5& 1& $1.2612\times 10^{-15}$& $2.8248\times 10^{-14}$\\ 
5& 2& $2.7925\times 10^{-12}$& $6.2543\times 10^{-11}$\\ 
5& 3& $1.0948\times 10^{-9}$& $2.4520\times 10^{-8}$\\ 
5& 4& $1.2334\times 10^{-8}$& $2.7625\times 10^{-7}$\\ 
5& 5& $9.4660\times 10^{-7}$& $2.1201\times 10^{-5}$\\ 
6& 1& $2.9141\times 10^{-19}$& $6.4421\times 10^{-18}$\\ 
6& 2& $1.3368\times 10^{-14}$& $2.9940\times 10^{-13}$\\ 
6& 3& $1.9677\times 10^{-12}$& $4.4070\times 10^{-11}$\\ 
6& 4& $3.5023\times 10^{-10}$& $7.8442\times 10^{-9}$\\ 
6& 5& $2.5728\times 10^{-9}$& $5.7623\times 10^{-8}$\\ 
6& 6& $1.9621\times 10^{-7}$& $4.3944\times 10^{-6}$\\ 
\hline
7& 1& $\dots                 $& $\dots$\\ 
7& 2& $9.2734\times 10^{-18}$& $2.0765\times 10^{-16}$\\ 
7& 3& $1.7446\times 10^{-14}$& $3.9073\times 10^{-13}$\\ 
7& 4& $8.2034\times 10^{-13}$& $1.8373\times 10^{-11}$\\ 
7& 5& $9.7500\times 10^{-11}$& $2.1837\times 10^{-9}$\\ 
7& 6& $5.3226\times 10^{-10}$& $1.1921\times 10^{-8}$\\ 
7& 7& $4.1412\times 10^{-8}$& $9.2750\times 10^{-7}$\\ 
\hline
8& 1& $\dots$                & $\dots$\\ 
8& 2& $2.8445\times 10^{-20}$ & $5.7900\times 10^{-19}$\\ 
8& 3& $2.1027\times 10^{-17}$ & $4.7089\times 10^{-16}$\\ 
8& 4& $1.0914\times 10^{-14}$ & $2.4445\times 10^{-13}$\\ 
8& 5& $2.6777\times 10^{-13}$ & $5.9973\times 10^{-12}$\\ 
8& 6& $2.5186\times 10^{-11}$ & $5.6409\times 10^{-10}$\\ 
8& 7& $1.0979\times 10^{-10}$ & $2.4590\times 10^{-9}$\\ 
8& 8& $8.8253\times 10^{-9}$  & $1.9766\times 10^{-7}$\\ 
\hline
\end{tabular}
\end{ruledtabular}
\end{center}
\end{table}

\begin{figure}[t]
\center
\includegraphics[width=0.46\textwidth]{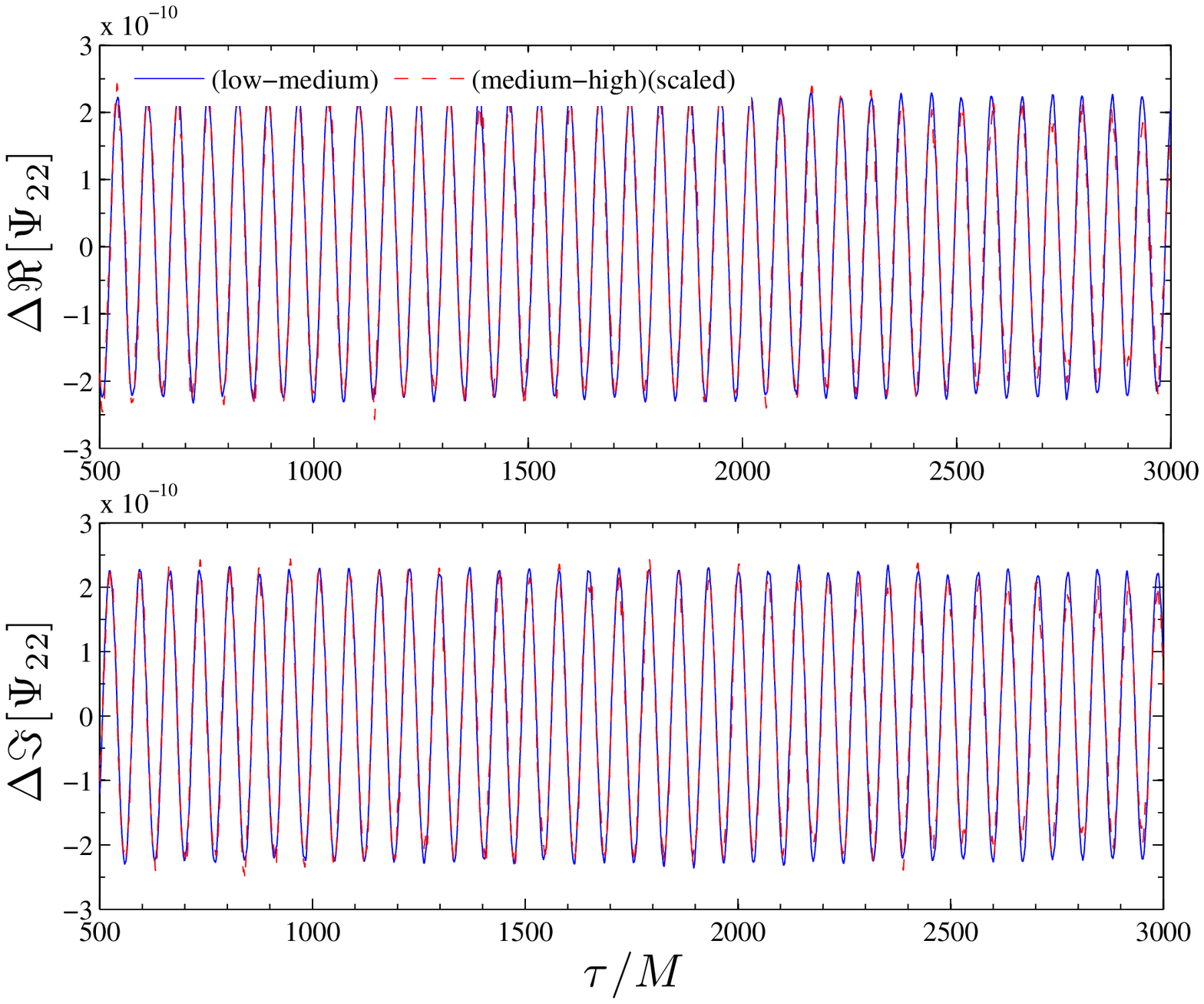}
\caption{\label{fig:ciro_conv} Convergence of real (top) and
  imaginary (bottom) part of the $\ell=m=2$ waveform generated by a
  particle on a circular orbit at $r_0=7.9456$. The plots show the 
  differences between low and medium resolution data and the difference 
  between medium and high resolution data scaled for 4th order convergence.}
\end{figure}

In this Appendix we present the convergence tests and an 
estimate of the errors on our data to validate our numerical approach. 
We observe the expected 4th order convergence, but 
also a progressive degradation of the quality of data relative
to sub-dominant multipoles with $\ell>6$ and decreasing index
$m\ll\ell$. 
In the following we consider the domain $[\rho_{\rm min},S]_{R_*}=[-50,70]_{50}$
discretized with $N=\{3001,\,6001,\,12001\}$ points, that correspond to
low, medium and high resolution. 
All data discussed in the bulk of the paper were obtained using the high resolution. 
The Courant factor is $\alpha_{\rm cfl}=0.5$, 
the Kreiss--Oliger dissipation factor is $\sigma=0.007$, 
the mass ratio considered for the tests is $\nu=10^{-2}$, 
and the waves are extracted at $\scri^+$. 
In the following plots we use the coordinate 
time $\tau$ on the horizontal axis.

We start by considering the waveforms emitted by a particle in stable
circular orbits (no radiation reaction) at $r_0=7.9456$. This value
of the radius is chosen here because it allows for an immediate comparison 
with published information~\cite{Martel:2003jj,Sopuerta:2005gz}.
Figure~\ref{fig:ciro_conv} shows the differences between low and medium
resolution data and the difference between medium and high resolution
data scaled for 4th order convergence (scaling factor, $s=16$) of the real (top panel) and the 
imaginary (bottom panel) part of the $\ell=m=2$ waveform. The differences
are superposed, thus indicating that the code converges at 
the correct rate. 
The plot does not show the initial junk radiation, but only the part
of the wave that is used below to calculate the GW energy flux. 
The behavior remains the same for all the subdominant multipoles. 
For multipoles $\ell\geq4$ and $m\to1$ however the amplitude of the 
solution becomes smaller and smaller (e.g.~$|\Psi_{61}|=A_{61}\simeq10^{-11}$),
until it becomes at the same order as round-off numerical noise, and
thus cannot be disentangled from it. In addition, such small-amplitude 
waves can also be polluted by reflections of the initial junk radiation 
(that remain always of the same order of magnitude for each multipole) 
from the internal boundary. 
In order to obtain cleaner data and to estimate the convergence rate in
these cases, we smooth the corresponding waves with a digital polynomial filter. 
The measurement of the convergence rate, however, becomes progressively more difficult 
for $\ell\geq4$ and, even with the smoothing, the $(8,1)$ and $(7,1)$ modes 
are completely polluted by high-frequency noise. Increasing the artificial 
dissipation in the code does not improve the results. In the future we shall 
investigate the possibility of reducing the initial junk radiation by 
improving the initial data set up. 
We shall also consider the use of  higher-order differential operators.

We recall, however, that the higher $\ell$ modes are progressively 
less relevant for the total waveform.

\begin{figure}[t]
\center
\includegraphics[width=0.46\textwidth]{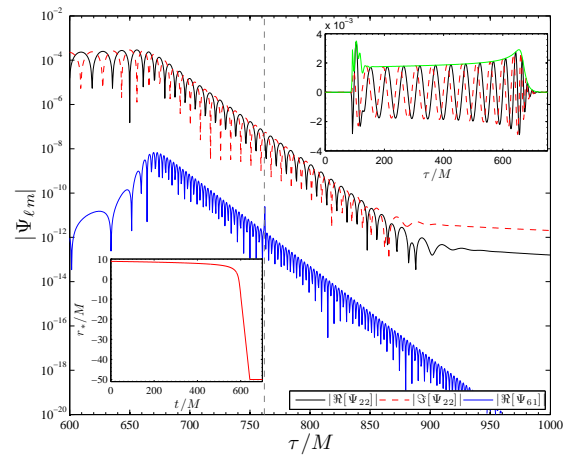}
\caption{\label{fig:inspl_mue2} (Color online) Multipolar waveforms generated by
the quasi-circular inspiral, plunge, merger and ringdown of the $\nu=10^{-2}$ binary 
initially at separation $r_0=7$. Main panel: the ringdown phase for the $\ell=m=2$ 
and $\ell=6$, $m=1$ modes. Top-right inset: the complete $\ell=m=2$ waveform. 
Bottom-left inset: the time-evolution of the $r_*$ coordinates of the point particle.
The vertical dashed line on the main panel marks the hyperboloidal time 
$\tau_{\rm end}\approx 762$ corresponding to the dynamical time $t$ where 
the particle reaches the  internal boundary of the numerical grid, 
$\rho_{\rm min}=-50$.}
\end{figure}

We finally list in Table~\ref{tab:gwflux} the values of the GW energy and
angular momentum fluxes emitted at $r_0=7.9456$, to be compared with 
published data~\cite{Martel:2003jj,Sopuerta:2005gz}.
The differences with the spectral data of Fujita et al.~\cite{Fujita:2004rb,Fujita:2009us} 
are below $0.8$~\% for each multipoles except for the multipoles $(7,7)$
($2.2$~\%), $(8,7)$ ($2.1$~\%) and $(8,8)$ ($4.8$~\%). 
We omit the values for multipoles $(7,1)$ and $(8,1)$ since they are not reliable.

\begin{figure}[t]
\center
\includegraphics[width=0.46\textwidth]{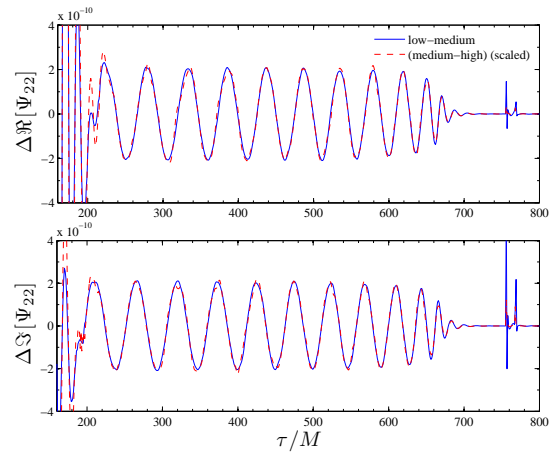}
\caption{\label{fig:inspl_conv} Same as Fig.~\ref{fig:ciro_conv} but
  for a particle in a inspiral to plunge orbit.}
\end{figure}

Now we discuss some general features of the multipolar waveforms
through transition from quasi-circular inspiral to plunge, merger 
and ringdown. The particle is initially at $r_0=7$.
The complete $\ell=m=2$ waveform, real (solid line) and imaginary 
part (dashed line) is displayed in the top-right inset of Fig.~\ref{fig:inspl_mue2}.
The main panel highlights the structure of the ringdown for the $\ell=m=2$ and the
$\ell=6$, $m=1$ modes. In the bottom-left inset of the figure we show the 
time-evolution of the radial $r_*$ coordinate of the particle: It is initially
at $r_*=8.8326$ and ends at $r_*=\rho_{\rm min}=-50$. 
When the  particle gets to $\rho_{\rm min}$ it is advected out of the grid, so
that the RWZ source becomes zero for the rest of evolution~\cite{Bernuzzi:2010ty,Damour:2007xr,Nagar:2006xv}. 
This jump in the source can introduce some artifacts in the ringdown waveform 
and thus the location of $\rho_{\rm min}$  should be chosen to minimize these
effects.
From the left-bottom inset of Fig.~\ref{fig:inspl_mue2} one sees that
$r_*=-50$ at $t_{\rm exit}\approx 642$. Since the speed of outgoing characteristics
is 1 on the layer by construction (see Eq.~\eqref{eq:speeds}), a signal
generated at $\rho_{\rm min}=-50$ will take a time $70-(-50)=120$ 
to reach $\scri^+$. This means that any signal connected with the particle
exiting the domain will show up on the waveform at $\scri^+$ at hyperboloidal
time $\tau_{\rm exit}=t_{\rm exit}+120=762$. This time is marked by the
vertical dashed line in Fig.~\ref{fig:inspl_mue2}. There is no
evidence of pathological features in the $\ell=m=2$ ringdown, but a localized
spike is seen in the (much smaller amplitude) $\ell=6$, $m=1$ multipole exactly at 
$\tau=\tau_{\rm exit}$. By inspecting all multipoles, we found
that the effect is always present when the waveform amplitude becomes
sufficiently small, e.g. for the $m\to 1 $ multipoles. Evidently, 
decreasing $\rho_{\rm min}$ (e.g., $\rho_{\rm min}=-200$) would delay
the occurrence of this spike, but not remove it,
because it is connected to our treatment of the particle and the coordinates
that we use (this problem should be solved in horizon-penetrating coordinates). 
The choice of $\rho_{\min}=-50$ is a reasonable compromise between 
efficiency and accuracy. 

We computed the decay rate of the tail of the
$\ell=m=2$ waveform. A linear fit to the initial part of the tail visible in 
Fig.~\ref{fig:inspl_mue2} gave around $-4.5$. This finding is in agreement
with Fig.~6 of~\cite{Zenginoglu:2009ey}, which solved the homogeneous RWZ in 
quadruple precision and 8th order finite differencing. The tail decay rate is expected
to approach the theoretical value $-(\ell+2)=-4$ asymptotically in time.

\begin{figure}[t]
\center
\includegraphics[width=0.46\textwidth]{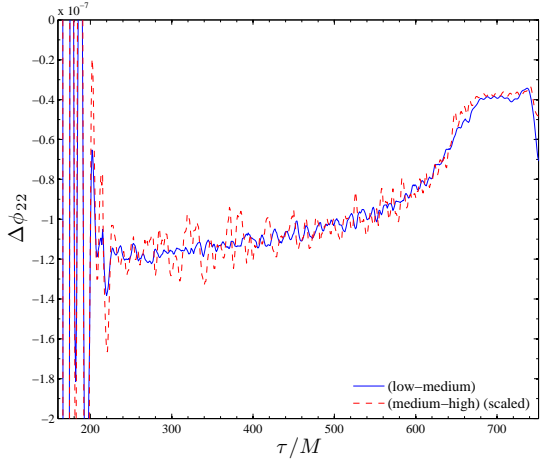}
\caption{\label{fig:inspl_conv_phi} Convergence of the phase 
  of the $\ell=m=2$ waveform for inspiral to plunge orbit.
  The plot shows the differences between low and medium 
  resolution data and the difference between medium and high
  resolution data scaled for 4th order convergence.}
\end{figure} 

The convergence of the $\ell=m=2$ waveform is demonstrated in 
Fig.~\ref{fig:inspl_conv}, that is the analogue of Fig.~\ref{fig:ciro_conv}. 
As expected, 4th order convergence is observed through inspiral, plunge, 
and merger phases, as well as during a considerable part of the ringdown. 
After time $\tau\approx 760$, one can notice the boundary effect mentioned above.

For the discussion in Sec.~\ref{sbsc:finitextr} it is important to establish
an error estimate for the gravitational wave phase.
Figure~\ref{fig:inspl_conv_phi} shows the convergence test on 
this quantity. The difference between the low and medium resolution
is around $\Delta\phi_{22}\lesssim10^{-7}$. 
Using Richardson extrapolation in resolution we estimate the error bars as
$\delta\phi_{22}\sim 10^{-6}$ and $\delta A_{22}/A_{22}\sim 10^{-6}$. 
We found similar results also for the other multipoles, when
possible, to establish the convergence rate.
As in the case of circular orbits, results are subjected to
a progressive degradation for sub-dominant modes $\ell\geq6$.

\section{Asymptotic formulas}
\label{app:asymp}

In this appendix we summarize the relations between the RWZ master
functions and the asymptotic observable quantities. From $\Psi^{(\rm e/o)}_{\ell
  m}$, the $h_+$ and $h_\times$ GW polarizations are obtained as
\be
\label{eq:hplus_cross}
\R \left(h_+-{\rm i}h_{\times}\right) = \sum_{\ell\geq 2,m}^{\ell_{\rm max}}\sqrt{\frac{(\ell+2)!}{(\ell-2)!}}
     \left(\Psi^{(\rm e)}_{\ell m}
     +{\rm i}\Psi^{(\rm o)}_{\ell m}\right)
     \;_{-2}Y^{\ell m} \ ,
\ee
where $\R$ is the distance from the source, $\ell_{\rm max}$ is the
maximum number of multipoles one sums over (omitted for brevity in the following sums) , and 
$\;_{-2}Y^{\ell m}\equiv\,_{-2}Y^{\ell m}(\theta,\varphi)$ are 
the $s=-2$ spin-weighted spherical harmonics. All second-order quantities
follow. The emitted power,
\be
\label{eq:dEdt}
\dot{E}  = \frac{1}{16\pi}\sum_{\ell\geq 2,m}\frac{(\ell+2)!}{(\ell-2)!}
	\left(\left|\dot{\Psi}^{(\rm o)}_{\ell m}\right|^2 + 
        \left|\dot{\Psi}^{(\rm e)}_{\ell m}\right|^2\right)\;, \\
\ee
the angular momentum flux
\be
\label{eq:dJdt}
\dot{J}  =  -\frac{1}{8\pi}\sum_{\ell\geq 2,m>0}
m\frac{(\ell+2)!}{(\ell -2)!}
\Im\left[\dot{\Psi}^{(\rm e)}_{\ell m}\Psi^{(\rm e)*}_{\ell m}
+\dot{\Psi}_{\ell m}^{({\rm o})}\Psi^{(\rm o)*}_{\ell m}\right], \\
\ee
that is obtained from the corresponding relation of Paper~I using 
$\Psi_{\lm}^*=(-1)^m\Psi_{\ell,-m}$, so that the sum is performed only
over $0<m\leq \ell$ multipoles,
and the linear momentum flux~\cite{Thorne:1980ru,Pollney:2007ss,Ruiz:2007yx},
\begin{align}
\label{eq:dPdt}
\F^{\bf P}_x + \ii\F^{\bf P}_y &= \dfrac{1}{8\pi}\sum_{\ell\geq 2,m}
\bigg[\ii a_{\lm} \dot{\Psi}_{\lm}^{({\rm e})}\dot{\Psi}^{({\rm o})*}_{\ell,m+1}\nonumber\\
&+b_{\lm}\left(\dot{\Psi}^{(\rm e)}_{\lm}\dot{\Psi}^{(\rm
    e)*}_{\ell+1,m+1}+\dot{\Psi}^{(\rm o)}_{\lm}\dot{\Psi}^{(\rm o)*}_{\ell
    +1,m+1}\right)\bigg] \ ,
\end{align}
where
\begin{align}
a_{\lm}    & = 2(\ell-1)(\ell+2)\sqrt{(\ell-m)(\ell+m+1)},\\
b_{\ell m} & =\dfrac{(\ell +3)!}{(\ell +1)(\ell -2)!}\sqrt{\dfrac{(\ell + m
    +1)(\ell+m+2)}{(2\ell+1)(2\ell +3)}} .
\end{align}

\bibliography{refs20111012}

\end{document}